\documentclass[12pt]{article}

\usepackage{amsmath,amssymb,amsfonts,amsthm}
\usepackage{graphicx}
\usepackage{cite}
\usepackage[all]{xy}
\usepackage[toc,page]{appendix}

\allowdisplaybreaks[3]

\newcounter{propositiona}
\newcommand{\propositiona}[1]{\refstepcounter{propositiona}
\noindent
\textbf{Proposition \thepropositiona.}\, {\it #1}}
\newcounter{definitiona}
\newcommand{\definitiona}[1]{\refstepcounter{definitiona}
\noindent
\textbf{Definition \thedefinitiona.}\, #1}
\newcounter{remarka}
\newcommand{\remarka}[1]{\refstepcounter{remarka}
\noindent
\textbf{Remark \theremarka.}\, #1}
\newcounter{examplea}
\newcommand{\examplea}[1]{\refstepcounter{examplea}
\noindent
\textbf{Example \theexamplea.}\, #1}
\newcounter{lemmaa}

\newcounter{theorema}
\newcommand{\theorema}[1]{\refstepcounter{theorema}
\noindent
\textbf{Theorem\, \thetheorema.}\, {\it #1}}
\newcounter{corollarya}

\begin{document}
	

\thispagestyle{empty}

\begin{center}
{\bf \Large Internal Lagrangians of PDEs as variational principles}\\[4ex]
{\large \bf Kostya Druzhkov}\\[2ex]
Russian-Armenian University, 123 Hovsep Emin street, Yerevan, 0051 Armenia\\
\textit{E-mail: konstantin.druzhkov@gmail.com}
\end{center}

\vspace{0.65ex}

\begin{abstract}\normalsize
A description of how the principle of stationary action reproduces itself in terms of the intrinsic geometry of variational equations is proposed. A notion of stationary points of an internal Lagrangian is introduced. A connection between symmetries, conservation laws and internal Lagrangians is established. Noether's theorem is formulated in terms of internal Lagrangians. A relation between non-degenerate Lagrangians and the corresponding internal Lagrangians is investigated. Several examples are discussed.\\[-1ex]
\end{abstract}


\section{Introduction}{\label{Intr}}

Many equations of mathematical physics originate from the principle of stationary action. Some equations are indirectly related to action functionals through their differential consequences. One can say that all such equations, in one way or another, have a variational nature. It seems reasonable to say that such equations encode their variational nature through elements of certain cohomologies, which we call internal Lagrangians~\cite{Druzhkov1}.

In this paper, we show that internal Lagrangians determine variational principles in terms of the intrinsic geometry of differential equations. Internal variational principles are related to integral functionals defined on particular classes of submanifolds of differential equations. Informally speaking, such submanifolds are composed of initial-boundary conditions lifted to infinitely prolonged equations. In this sense, one can say that such internal variational principles are related to the covariant phase space formalism (see, e.g.,~\cite{CrnWit, Got1, GotIsMa, Helein, Khavk1, IbSp} and references therein). However, internal Lagrangians belong to the intrinsic geometry of infinitely prolonged differential equations. 
Despite the generality of the considered construction, its application to gauge theories proves to be challenging. The approach needs some modification in this case.

In a broad sense, the paper is related to the inverse problem of the calculus of variations (see, e.g.,~\cite{Henn, Krup, Khavk2, Grigoriev, Druzhkov2, GriGri} and references therein). Basically, we are interested in how variational principles originating from action functionals (on jets) manifest themselves at the level of the intrinsic geometry of differential equations.

The paper is organized as follows. In Section~\ref{Def} we recall some basic concepts from the geometry of differential equations. 
Section~\ref{VariL} proposes an explanation of why any stationary action principle for a system of differential equations is encoded in its intrinsic geometry. A notion of stationary points of an internal Lagrangian is introduced.
Section~\ref{ILcl} focuses on a relation between internal Lagrangians and conservation laws.
In Section~\ref{iLs} we present Noether's theorem in terms of the intrinsic geometry of differential equations and discuss gauge invariance of internal Lagrangians. Notably, this version of Noether's theorem applies to all symmetries, not only to variational ones~\cite{Olver}. In a nutshell, any symmetry of a variational system of differential equations either gives rise to conservation laws or generates a nontrivial internal Lagrangian of the system.
Section~\ref{Compl} is devoted to the Euler-Lagrange equations for non-degenerate Lagrangians. We view differential equations as bundles and describe another variational principle that takes into account space+time decompositions of the base spaces. 
Section~\ref{Exam} contains several simple examples. Section~\ref{degen} discusses the Proca equations as an example of the Euler-Lagrange equations for a degenerate Lagrangian.

\textit{All manifolds and functions considered in this paper are assumed to be smooth of the class $C^{\infty}$. All submanifolds are assumed to be embedded.}

\section{Jets and equations}\label{Def}

Let us introduce some notation and briefly recall basic facts from the geometry of differential equations. More details can be found in~\cite{VinKr, KraVer1}.

\subsection{Jets}

Let $\pi\colon E\to M$ be a locally trivial smooth vector
bundle over a smooth manifold $M$, $\mathrm{dim}\, M = n$, 
$\mathrm{dim}\, E = n + m$. The bundle $\pi$ allows one to introduce the corresponding jet bundles
\begin{align*}
\xymatrix{
\ldots \ar[r] & J^3(\pi) \ar[r]^-{\pi_{3, 2}} & J^2(\pi) \ar[r]^-{\pi_{2, 1}} & J^1(\pi) \ar[r]^-{\pi_{1, 0}} & J^0(\pi) = E \ar[r]^-\pi & M
}
\end{align*}
and the inverse limit $J^{\infty}(\pi)$ of the chain. The manifold $J^{\infty}(\pi)$ arises together with the natural projections 
$\pi_{\infty}\colon J^{\infty}(\pi)\to M$ and $\pi_{\infty,\, k}\colon J^{\infty}(\pi)\to J^k(\pi)$. Denote by $\mathcal{F}(\pi)$ the algebra of smooth functions on~$J^{\infty}(\pi)$.\\[-2ex]

\noindent
\textbf{Local coordinates.} Suppose $U\subset M$ is a coordinate neighborhood such that the bundle $\pi$
becomes trivial over $U$. Choose local coordinates $x^1$, \ldots, $x^n$ in $U$ and $u^1$, \ldots, $u^m$  
along the fibers of $\pi$ over $U$. It is convenient to introduce a
multi-index $\alpha$ as a formal sum of the form $\alpha = \alpha_1 x^1 + \ldots + \alpha_n x^n = \alpha_i x^i$, where all $\alpha_i$ are non-negative integers; $|\alpha| = \alpha_1 + \ldots + \alpha_n$.
We denote by $u^i_{\alpha}$ the corresponding adapted local coordinates on $J^{\infty}(\pi)$.\\[-2ex]

\noindent
\textbf{Cartan distribution.} The main structure on $J^{\infty}(\pi)$ is the Cartan distribution $\mathcal{C}$.
The Cartan distribution is spanned by the total derivatives (using the summation convention)
$$
D_{x^i} = \partial_{x^i} + u^k_{\alpha + x^i}\partial_{u^k_{\alpha}}\qquad\quad i = 1,\ldots, n.
$$
Historically, this distribution is also associated with Johann Friedrich Pfaff, Sophus Lie, etc.\\[-2ex]

\noindent
\textbf{Cartan forms.} The Cartan distribution determines the ideal $\mathcal{C}\Lambda^*(\pi)\subset \Lambda^*(\pi)$ 
of the algebra of differential forms on $J^{\infty}(\pi)$.
The ideal $\mathcal{C}\Lambda^*(\pi)$ is generated by Cartan forms, i.e., differential forms vanishing on the Cartan distribution.
A Cartan $1$-form $\omega\in\mathcal{C}\Lambda^1(\pi)$ can be written as a finite sum
$$
\omega = \omega_i^{\alpha}\theta^i_{\alpha}\,,\qquad\ \theta^i_{\alpha} = du^i_{\alpha} - u^i_{\alpha + x^k}dx^k
$$
in adapted local coordinates. Here the coefficients $\omega_i^{\alpha}$ are smooth functions defined on a coordinate domain of $J^{\infty}(\pi)$. We denote by $\mathcal{C}^p\Lambda^*(\pi)$ the $p$-th power of the ideal $\mathcal{C}\Lambda^*(\pi)$.\\[-2ex]

\noindent
\textbf{Infinitesimal symmetries.}
Let $\varkappa(\pi) = \Gamma(\pi^*_{\infty}(\pi))$ be the $\mathcal{F}(\pi)$-module
of sections of the pullback $\pi^*_{\infty}(\pi)$. 
If $\varphi\in \varkappa(\pi)$, there is the corresponding evolutionary vector field on $J^{\infty}(\pi)$
$$
E_{\varphi} = D_{\alpha}(\varphi^i)\partial_{u^i_{\alpha}}\,,
$$
where $\varphi^1$, \ldots, $\varphi^m$ are components of $\varphi$ in adapted local coordinates, $D_{\alpha}$ denotes the composition $D_{x^1}^{\ \alpha_1}\circ\ldots\circ D_{x^n}^{\ \alpha_n}$. Evolutionary vector fields are infinitesimal symmetries of $J^{\infty}(\pi)$. In particular, $\mathcal{L}_{E_{\varphi}}(\mathcal{C}\Lambda^*(\pi))\subset \mathcal{C}\Lambda^*(\pi)$. Here $\mathcal{L}_{E_{\varphi}}$ is the corresponding Lie derivative.\\[-2ex]

\noindent
\textbf{Horizontal forms.}
Cartan forms allow one to consider horizontal $k$-forms
$$
\Lambda^k_h(\pi) = \Lambda^k(\pi)/\mathcal{C}\Lambda^k(\pi)\,.
$$
A differential $k$-form $\omega\in\Lambda^k(\pi)$ represents the corresponding horizontal $k$-form $[\omega]_h = \omega + \mathcal{C}\Lambda^k(\pi)$.
The infinite jet bundle $\pi_{\infty}\colon J^{\infty}(\pi) \to M$ admits the decomposition 
$$
\Lambda^1(\pi) = \mathcal{C}\Lambda^1(\pi) \oplus \mathcal{F}(\pi)\!\cdot\!\pi^*_{\infty}(\Lambda^1(M))\,.
$$
The module of horizontal $k$-forms $\Lambda^k_h(\pi)$ can be identified with the module $\mathcal{F}(\pi)\cdot \pi^*_{\infty}(\Lambda^k(M))$.
According to the inclusion $d(\mathcal{C}\Lambda^*(\pi)) \subset \mathcal{C}\Lambda^*(\pi)$, the de Rham differential $d$ induces the horizontal differential
\begin{equation*}
d_h\colon \Lambda^{*}_h(\pi)\to \Lambda^{*}_h(\pi).
\end{equation*}

\noindent
\textbf{Adjoint modules.}
Let $\eta\colon E_1\to M$ be a locally trivial smooth vector
bundle. Suppose $P(\pi)$ is the module of sections of the pullback $\pi^*_{\infty}(\eta)$:
$$
P(\pi) = \Gamma(\pi^*_{\infty}(\eta))\,.
$$
There exists the adjoint module $\widehat{P}(\pi)$,
\begin{equation*}
\widehat{P}(\pi) = \mathrm{Hom}_{\mathcal{F}(\pi)}(P(\pi), \Lambda^n_h(\pi))\,.
\end{equation*}
\noindent
\textbf{Euler operator.}
By $\mathrm{E}$ we denote the Euler operator (variational derivative), $\mathrm{E}\colon \Lambda^n_h(\pi)\to \widehat{\varkappa}(\pi)$. Suppose $L\in\Lambda^n(\pi)$ is a differential $n$-form. If $L$ is of the form $\lambda \, dx^1\wedge\ldots\wedge dx^n$, then
\begin{align*}
&\mathrm{E}[L]_h = (-1)^{|\alpha|}D_{\alpha}\Big(\dfrac{\partial \lambda}{\partial u^i_{\alpha}}\Big)\, \theta^i_0\wedge dx^1\wedge\ldots\wedge dx^n\,,\\
&\langle \mathrm{E}[L]_h, \varphi\rangle = i_{E_{\varphi}}\,\mathrm{E}[L]_h = (-1)^{|\alpha|}D_{\alpha}\Big(\dfrac{\partial \lambda}{\partial u^i_{\alpha}}\Big)\varphi^i\, dx^1\wedge\ldots\wedge dx^n\,.
\end{align*}
Here $\langle \cdot , \cdot \rangle$ is the natural pairing between a module and its adjoint.

The Noether formula links the Lie derivative $\mathcal{L}_{E_{\varphi}}[L]_h$ of a horizontal $n$-form $[L]_h$ and the variational derivative $\mathrm{E}[L]_h$. Namely,
there exists a Cartan form $\omega_L\in \mathcal{C}\Lambda^{n}(\pi)$ such that
\begin{align}
\mathcal{L}_{E_{\varphi}}[L]_h = \langle \mathrm{E}[L]_h, \varphi \rangle + d_h [i_{E_{\varphi}}\omega_L]_h\,.
\label{Noeth}
\end{align}
By $\delta/\delta u^i$ we denote the variational derivative (or simply variation) with respect to the variable~$u^i$,
\begin{align*}
\dfrac{\delta\, }{\delta u^i} = (-1)^{|\alpha|}D_{\alpha}\circ \partial_{u^i_{\alpha}}\,.
\end{align*}

\subsection{Differential equations \label{difeq}}

Let $F$ be a (smooth) section of some bundle of the form $\pi^*_{r}(\eta)$, where $\pi_r\colon J^r(\pi)\to M$. We assume that, for each point $p\in \{F = 0\}\subset J^{r}(\pi)$, the differentials of coordinate functions $dF^i|_{p}$ are linearly independent. So, $\{F = 0\}\subset J^{r}(\pi)$ is an embedded submanifold. By the infinite prolongation of the differential equation $F = 0$ we mean the subset $\mathcal{E}\subset J^{\infty}(\pi)$ defined by the following infinite system of equations
\begin{align*}
\quad D_{\alpha}(F^i) = 0\,,\qquad |\alpha| \geqslant 0\,.
\end{align*}
Henceforth, we assume that $\pi_{\infty}(\mathcal{E}) = M$.\\[-1ex]

\remarka{We do not require that the number of equations of the form $F^i = 0$ coincide with the number of unknowns. We also do not require a bundle structure on a differential equation (except in Section~\ref{Compl} and examples).}\\[-1 ex]

By $\mathcal{F}(\mathcal{E})$ we denote the algebra of smooth functions on $\mathcal{E}$, 
$$
\mathcal{F}(\mathcal{E}) = \mathcal{F}(\pi)|_{\mathcal{E}} = \mathcal{F}(\pi)/I\,.
$$
Here $I$ denotes the ideal of the system $\mathcal{E}\subset J^{\infty}(\pi)$. The algebra of smooth functions produces the algebra of differential forms $\Lambda^*(\mathcal{E}) = \Lambda^*(\pi)|_{\mathcal{E}}$. The Cartan distribution on $J^{\infty}(\pi)$ can be restricted to $\mathcal{E}$. We denote by $\mathcal{C}_p$ the Cartan plane at a point $p\in\mathcal{E}$. 
Similarly, there is the ideal $\mathcal{C}\Lambda^*(\mathcal{E})\subset \Lambda^*(\mathcal{E})$ consisting of differential forms that vanish on the Cartan distribution (on $\mathcal{E}$). Let us also introduce the following $\mathcal{F}(\mathcal{E})$-modules
$$
\varkappa(\mathcal{E}) = \varkappa(\pi)/I\cdot \varkappa(\pi)\,,\qquad\quad P(\mathcal{E}) = P(\pi)/I\cdot P(\pi).
$$

\noindent
\textbf{$\mathcal{C}$-differential operators.} By \textit{$\mathcal{C}$-differential operators} we mean operators in total derivatives. If $Q_1$ and $Q_2$ are modules of sections of some vector bundles over an infinitely prolonged system of differential equations $\mathcal{E}$, we denote by $\mathcal{C}(Q_1, Q_2)$ the module of $\mathcal{C}$-differential operators from $Q_1$ to $Q_2$.\\[-2ex]

\noindent
\textbf{Regularity assumptions.}
We say that the infinite prolongation $\mathcal{E}$ of a system of differential equations $F = 0$ is \textit{regular} if for each
function $f\in \mathcal{F}(\pi)$ vanishing on $\mathcal{E}$ (i.e., $f\in I$), there exists a $\mathcal{C}$-differential operator $\Delta\colon P(\pi)\to \mathcal{F}(\pi)$ such that $f = \Delta(F)$.

\textit{In what follows, we
assume that the de Rham cohomology groups $H^i_{dR}(\mathcal{E})$ are trivial for $i > 0$ and consider only regular systems.} These conditions are not restrictive.
The case of non-trivial de Rham cohomology does not lead to essential complications. In this case, one can consider quotients by
subgroups of topological (locally trivial) elements.\\[-2ex]

\noindent
\textbf{Infinitesimal symmetries.} A \textit{symmetry} (more precisely, infinitesimal symmetry) of an infinitely prolonged system of equations $\mathcal{E}$ is a vector field $X\in D(\mathcal{E})$ preserving the Cartan distribution in the following sense
$$
[X, \mathcal{C}D(\mathcal{E})]\subset \mathcal{C}D(\mathcal{E}).
$$
Here $\mathcal{C}D(\mathcal{E})\subset D(\mathcal{E})$ is the module of Cartan vector fields, i.e., linear combinations of total derivatives, with coefficients in $\mathcal{F}(\mathcal{E})$.
Elements of $\mathcal{C}D(\mathcal{E})$ are trivial symmetries. Two symmetries of $\mathcal{E}$ are equivalent if they differ by a trivial one.

A differential equation $\{F = 0\}\subset J^{r}(\pi)$ allows us to introduce the linearization $l_{\mathcal{E}}\colon \varkappa(\mathcal{E})\to P(\mathcal{E})$. To this end, we define the $\mathcal{C}$-differential operator $l_F\colon \varkappa(\pi)\to P(\pi)$ by $l_F(\varphi) = E_{\varphi}(F)$ and set $l_{\mathcal{E}} = l_F|_{\mathcal{E}}$.

If $\pi_{\infty,\, 0}(\mathcal{E}) = J^0(\pi)$, symmetries of $\mathcal{E}$ can be identified with elements of $\ker l_{\mathcal{E}}$ using their characteristics (see, e.g.,~\cite{Olver},~\cite{VinKr}).
A gauge symmetry of $\mathcal{E}$ is a symmetry of the form $R(\epsilon)\in \varkappa(\mathcal{E})$, where $R$ is a $\mathcal{C}$-differential operator such that $l_{\mathcal{E}}\circ R = 0$.\\[-2ex]

\noindent
\textbf{$\mathcal{C}$-spectral sequence.} Powers of the ideal $\mathcal{C}\Lambda^*(\mathcal{E})$ are stable with respect to the de Rham differential, i.e., $d(\mathcal{C}^p\Lambda^*(\mathcal{E}))\subset \mathcal{C}^p\Lambda^*(\mathcal{E})$, where
$$
\mathcal{C}^p\Lambda^*(\mathcal{E}) = \underbrace{\mathcal{C}\Lambda^1(\mathcal{E})\wedge\ldots\wedge \mathcal{C}\Lambda^1(\mathcal{E})}_{p}\wedge\, \Lambda^*(\mathcal{E})\,.
$$
Thus, the de Rham complex admits the filtration
$$
\Lambda^*(\mathcal{E})\supset \mathcal{C}\Lambda^*(\mathcal{E})\supset \mathcal{C}^2\Lambda^*(\mathcal{E})\supset \mathcal{C}^3\Lambda^*(\mathcal{E})\supset \ldots
$$
The corresponding spectral sequence $(E^{p,\, q}_r(\mathcal{E}), d_r)$ is the Vinogradov $\mathcal{C}$-spectral sequence~\cite{Vin}. 
In particular, the $\mathcal{C}$-spectral sequence allows one to define conservation laws, variational $1$-forms and presymplectic structures of differential equations.

A \textit{conservation law} of an infinitely prolonged system $\mathcal{E}$ is an element of the group (more precisely, of the vector space) $E^{\,0,\,n-1}_1(\mathcal{E})$. A \textit{variational $1$-form} of $\mathcal{E}$ is an element of the group $E^{\,1,\,n-1}_1(\mathcal{E})$. Conservation laws and variational $1$-forms are related by the differential
$$
d_1^{\,0,\,n-1}\colon E^{\,0,\,n-1}_1(\mathcal{E}) \to E^{1,\,n-1}_1(\mathcal{E}).
$$
A \textit{presymplectic structure} of $\mathcal{E}$ is an element of the kernel of the differential
$$
d_1^{\,2,\,n-1}\colon E^{\,2,\,n-1}_1(\mathcal{E})\to E^{\,3,\,n-1}_1(\mathcal{E}).
$$
A presymplectic structure of $\mathcal{E}$ can be (ambiguously) represented by a $\mathcal{C}$-differential operator $\Delta\colon \varkappa(\mathcal{E})\to \widehat{P}(\mathcal{E})$ such that $l^{\,*}_{\mathcal{E}}\circ \Delta = \Delta^* \circ l_{\mathcal{E}}$ (see, e.g.,~\cite{VinKr}).\\[-2ex]

\noindent
\textbf{Cosymmetries (see, e.g.,~\cite{VinKr}).} A \textit{cosymmetry} of an infinitely prolonged system $\mathcal{E}$ is an element of $\ker l^{\,*}_{\mathcal{E}}$. Cosymmetries are related to variational $1$-forms. Namely, if $\psi$ is a cosymmetry of $\mathcal{E}$, there is a $\mathcal{C}$-differential operator $\nabla\colon \varkappa(\mathcal{E})\to \Lambda^{n-1}_h(\mathcal{E})$ such that
\begin{align}
\langle \psi, l_{\mathcal{E}}(\varphi) \rangle = d_h\nabla(\varphi)\qquad \text{for all}\ \ \varphi\in \varkappa(\mathcal{E})\,.
\label{cos1f}
\end{align}
The operator $\nabla$ determines a $d_0$-closed element of the group $E^{1,\, n-1}_0(\mathcal{E})$. The corresponding variational $1$-form is well defined. In fact, this mapping from cosymmetries to variational $1$-forms is always surjective.

If an operator $\Delta$ represents a presymplectic structure $\omega\in \ker d_1^{\,2,\,n-1}$ and $\varphi\in \ker l_{\mathcal{E}}$ corresponds to a symmetry $X$, then $\Delta(\varphi)$ is a cosymmetry, and $\Delta(\varphi)$ corresponds to the variational $1$-form $i_X\omega$.\\[-2ex]

\noindent
\textbf{Internal Lagrangians (see~\cite{Druzhkov1}).} An \textit{internal Lagrangian} of an infinitely prolonged system $\mathcal{E}$ is an element of the group
\begin{align*}
\widetilde{E}^{\,0,\, n-1}_1(\mathcal{E}) = \dfrac{\{l\in\Lambda^n(\mathcal{E})\colon\ dl\in \mathcal{C}^2\Lambda^{n+1}(\mathcal{E})\}}
{\mathcal{C}^2\Lambda^{n}(\mathcal{E}) + d(\Lambda^{n-1}(\mathcal{E}))}\,.
\end{align*}

If the variational derivative $\mathrm{E}[L]_h$ of a 
horizontal $n$-form $[L]_h\in\Lambda^n_h(\pi)$ vanishes on $\mathcal{E}\subset J^{\infty}(\pi)$, then there exists a Cartan $n$-form $\omega_L\in\mathcal{C}\Lambda^{n}(\pi)$ such that $d(L + \omega_L) - \mathrm{E}[L]_h \in\mathcal{C}^2\Lambda^{n+1}(\pi)$. For example, one can take the generalized Poincar\'e-Cartan form (or any Lepage equivalent of $L$) as $L + \omega_L$, provided $L\in\mathcal{F}(\pi)\cdot \pi_{\infty}^*(\Lambda^n(M))$. The form $(L + \omega_L)|_{\mathcal{E}}$ represents an element of the group
\begin{align}
\dfrac{\{l\in\Lambda^n(\mathcal{E})\colon\ dl\in \mathcal{C}^2\Lambda^{n+1}(\mathcal{E})\}}
{\mathcal{C}^2\Lambda^{n}(\mathcal{E}) + d(\mathcal{C}\Lambda^{n-1}(\mathcal{E}))}
\label{SIL}
\end{align}
and the corresponding internal Lagrangian. If an infinitely prolonged system $\mathcal{E}$ is embedded into some jets $J^{\infty}(\pi)$, and $\mathcal{C}\Lambda^*(\mathcal{E}) = \mathcal{C}\Lambda^*(\pi)|_{\mathcal{E}}$, then each element of group~\eqref{SIL} ambiguously defines a (global) horizontal $n$-form $[L]_h\in\Lambda^n_h(\pi)$ such that $\mathrm{E}[L]_h|_{\mathcal{E}} = 0$ (see~\cite{Druzhkov1}, Theorem 1). In essence, this fact is based on the results obtained in~\cite{Khavk2} (Theorem 3).

The de Rham differential $d$ induces the differential
$$
\tilde{d}^{\,0,\,n-1}_1\colon \widetilde{E}^{\,0,\, n-1}_1(\mathcal{E})\to E^{\,2,\, n-1}_1(\mathcal{E})\,,
$$
which maps internal Lagrangians to presymplectic structures (i.e., $\mathrm{im}\, \tilde{d}^{\,0,\,n-1}_1\subset \ker d_1^{\,2,\,n-1}$).

\section{\label{VariL} Stationary points of internal Lagrangians}

Let $\mathcal{E}$ be an infinitely prolonged system of differential equations with $n$-dimensional Cartan distribution, and let $l\in \Lambda^n(\mathcal{E})$ be a differential form such that $dl\in\mathcal{C}^2\Lambda^{n+1}(\mathcal{E})$. Suppose $N$ is a compact $n$-dimensional oriented smooth manifold with boundary $\partial N$.
Our immediate goal is to relate the equivalence class 
$$
l + \mathcal{C}^2\Lambda^{n}(\mathcal{E}) + d(\mathcal{C}\Lambda^{n-1}(\mathcal{E}))
$$
to an integral functional.\\[-1ex]

\definitiona{An embedding $\sigma\colon N\to \mathcal{E}$ is an \textit{almost Cartan embedding} if
\begin{align*}
\dim\, (T_p \,\sigma(N) \cap \mathcal{C}_p) \geqslant n-1 \qquad \text{for all}\ \ p\in \sigma(N)\,.
\end{align*}
}
We denote by $\mathcal{A}_N(\mathcal{E})$ the set of almost Cartan embeddings of $N$ to $\mathcal{E}$. 
It is easy to see that $\sigma^*(\mathcal{C}^2\Lambda^{n}(\mathcal{E})) = 0$ for any $\sigma\in \mathcal{A}_N(\mathcal{E})$.\\[-1ex]

\examplea{Consider the heat equation and some initial conditions for all $t_0\in\mathbb{R}$
$$
u_t = u_{2x}\,,\qquad u = f(x, t_0)\,.
$$
For each fixed $t_0$, one can determine all the derivatives of $u$ using these data. Namely,
\begin{align}
\begin{aligned}
&u = f(x, t_0)\,,\qquad u_x = \partial_x f(x, t_0)\,,\qquad u_t = \partial_x^2 f(x, t_0)\,,\\
&u_{2x} = \partial_x^2 f(x, t_0)\,,\qquad u_{x+t} = \partial_x^3 f(x, t_0)\,,\qquad u_{2t} = \partial_x^{4} f(x, t_0)\,,\qquad\ldots
\end{aligned}
\label{heat}
\end{align}
Substituting $t$ for $t_0$ in~\eqref{heat}, we obtain an embedding of $\mathbb{R}^2$ to the corresponding infinitely prolonged system. The restriction of this embedding to a compact submanifold of $\mathbb{R}^2$ is an almost Cartan embedding since its differential maps vectors of the form $\partial_x$ to vectors of the form $\,\overline{\!D}_x = D_x|_{\mathcal{E}}$.\\[-1ex]
}

The \textit{images} of almost Cartan embeddings play a crucial role in the construction of integral functionals determined by the intrinsic geometry of differential equations. Since the Cartan distribution of $\mathcal{E}$ is involutive, one can say (informally) that such images are composed of initial-boundary conditions lifted to $\mathcal{E}$. However, speaking of stationary points of internal Lagrangians, we prefer the Lagrangian approach to the Eulerian one and consider perturbations of embeddings rather than motions of their images. 

One can say that almost Cartan embeddings of $N$ to $\mathcal{E}$ are related to spatial parts of space+time decompositions of $N$. More precisely, in the presence of $\sigma\in \mathcal{A}_N(\mathcal{E})$, a surface $\Sigma\subset N$ of codimension~$1$ can be considered a spatial one (at some instant) if $d\sigma(T_x\Sigma) \subset \mathcal{C}_{\sigma(x)}$ for all $x\in \Sigma$ (see also Section~\ref{Compl}).\\[-1ex]

\remarka{If $\mathcal{E}$ is an infinitely prolonged system of ODEs, then all embeddings of $N$ to $\mathcal{E}$ are almost Cartan embeddings.}\\[-1ex]

\remarka{An embedding $s\colon N\to \mathcal{E}$ defines a solution to $\mathcal{E}$ if and only if
\begin{align*}
\dim\, (T_p \,s(N) \cap \mathcal{C}_p) = n\qquad \text{for all}\quad p\in s(N)\,.
\end{align*}
}

\definitiona{Let $\sigma\colon N\to \mathcal{E}$ be an almost Cartan embedding. We say that $\sigma$ \textit{defines a boundary value problem} if
$T_p \, \sigma(\partial N) \subset \mathcal{C}_p$ for all $p\in \sigma(\partial N)$.
}\\[-1ex]

Denote by $\mathcal{BA}_N(\mathcal{E})$ the set of almost Cartan embeddings (of $N$ to $\mathcal{E}$) that define boundary value problems. 
If $\sigma\in \mathcal{BA}_N(\mathcal{E})$, then Stokes' theorem implies
\begin{align*}
\int_N\sigma^*(\omega) = 0\qquad \text{for all}\quad \omega\in d(\mathcal{C}\Lambda^{n-1}(\mathcal{E}))\,.
\end{align*}
So, the equivalence class $l + \mathcal{C}^2\Lambda^{n}(\mathcal{E}) + d(\mathcal{C}\Lambda^{n-1}(\mathcal{E}))$ does indeed define the integral functional
\begin{align}
S\colon \mathcal{BA}_N(\mathcal{E}) \to \mathbb{R}\,,\qquad S(\sigma) = \int_N \sigma^*(l) = \int_{\sigma(N)} l|_{\sigma(N)}\,.
\label{IntAct}
\end{align}
The functional $S$ is diffeomorphism-invariant in the following sense. If $f\colon N \to N$ is a diffeomorphism that preserves the orientation, and $\sigma\in \mathcal{BA}_N(\mathcal{E})$, then $\sigma \circ f\in \mathcal{BA}_N(\mathcal{E})$ and $S(\sigma \circ f) = S(\sigma)$.

We can consider derivatives of the functional $S$ along paths in $\mathcal{BA}_N(\mathcal{E})$. However, in this paper, we define stationary points of internal Lagrangians. Nevertheless, this functional clarifies why a variational nature of a system of differential equations is encoded in the intrinsic geometry of such a system: \textit{any stationary action principle for a system of differential equations gives rise to a variational principle in terms of its intrinsic geometry.}\\[-1ex]

\remarka{If the variational derivative $\mathrm{E}[L]_h$ of a 
horizontal $n$-form $[L]_h\in\Lambda^n_h(\pi)$ vanishes on an infinitely prolonged system $\mathcal{E}\subset J^{\infty}(\pi)$, then $[L]_h$ defines a unique functional of the form~\eqref{IntAct}. In addition, the cohomology class $[L]_h + d_h (\Lambda^{n-1}_h(\pi))$ defines a unique internal Lagrangian in this case.}\\[-1ex]

The form $l$ also represents the internal Lagrangian 
$$
\boldsymbol\ell = l + \mathcal{C}^2\Lambda^{n}(\mathcal{E}) + d(\Lambda^{n-1}(\mathcal{E}))\,.
$$
Because of non-trivial boundary terms, an internal Lagrangian of $\mathcal{E}$ ambiguously defines an integral functional on $\mathcal{A}_N(\mathcal{E})$. Nonetheless, its derivative along a path is well defined. Such a derivative is completely determined by the corresponding presymplectic structure.
To see this, consider a path in $\mathcal{A}_N({\mathcal{E}})$, i.e., a smooth mapping $\gamma\colon \mathbb{R}\times N\to \mathcal{E}$ such that for all $\tau\in\mathbb{R}$, the mappings
$$
\gamma(\tau)\colon N\to \mathcal{E},\qquad \gamma(\tau)\colon x\mapsto \gamma(\tau, x)
$$ 
are almost Cartan embeddings. Define $0_N\colon N\to \mathbb{R}\times N$ by $0_N(x) = (0, x)$. Then $\gamma(0) = \gamma\circ 0_N$, and the homotopy formula
\begin{align*}
\dfrac{d}{d\tau}\Big|_{\tau = 0}\gamma(\tau)^*(l) = 0_N^*\big(\mathcal{L}_{\partial_{\tau}}\gamma^*(l)\big) = d\,0_N^*\big(i_{\partial_{\tau}}\gamma^*(l)\big) + 0_N^*\big(i_{\partial_{\tau}}\gamma^*(dl)\big)
\end{align*}
holds. Suppose that each point of the boundary is fixed, i.e., for each $x\in \partial N$,
\begin{align}
\gamma(\tau)(x) = \gamma(0)(x)\qquad \text{for all} \quad \tau\in\mathbb{R}\,.
\label{boundfix}
\end{align}
Then the form $0_N^*\big(i_{\partial_{\tau}}\gamma^*(l)\big)$ vanishes on $\partial N$, and the derivative along $\gamma$
\begin{align*}
\dfrac{d}{d\tau}\Big|_{\tau = 0}\int_N \gamma(\tau)^*(l) = \int_N 0_N^*\big(i_{\partial_{\tau}}\gamma^*(dl)\big)
\end{align*}
is determined by the presymplectic structure represented by $dl$.\\[-1ex] 

\definitiona{An almost Cartan embedding $\sigma\colon N \to \mathcal{E}$ is a \textit{stationary point} of an internal Lagrangian $\boldsymbol\ell = l + \mathcal{C}^2\Lambda^{n}(\mathcal{E}) + d(\Lambda^{n-1}(\mathcal{E}))$ if
\begin{align*}
\dfrac{d}{d\tau}\Big|_{\tau = 0}\int_N \gamma(\tau)^*(l) = 0
\end{align*}
for any path $\gamma$ in $\mathcal{A}_N(\mathcal{E})$ such that $\gamma(0) = \sigma$ and each point of the boundary is fixed.}\\[-1ex]

\remarka{\label{boufix} Instead of~\eqref{boundfix}, one can use the following condition: $\gamma(\tau)(\partial N) = \gamma(0)(\partial N)$ for all $\tau\in\mathbb{R}$. In this case, the form $0_N^*\big(i_{\partial_{\tau}}\gamma^*(l)\big)$ also vanishes on $\partial N$.}\\[-1ex]

The form $dl$ belongs to $\mathcal{C}^2\Lambda^{n+1}(\mathcal{E})$. Then, substituting $n$ vectors from a Cartan plane to $dl$, we get zero. Hence,
if the embedding $\gamma(0)$ defines a solution of $\mathcal{E}$, then $0_N^*\big(i_{\partial_{\tau}}\gamma^*(dl)\big) = 0$. We obtain\\[-1ex]

\propositiona{\label{Sol} Let $\sigma\colon N \to \mathcal{E}$ be an embedding that defines a solution to an infinitely prolonged system of equations $\mathcal{E}$. Then $\sigma$ is a stationary point of any internal Lagrangian of $\mathcal{E}$.}\\[-1 ex]

Let us note that a result close in meaning to Proposition~\ref{Sol} is given in~\cite{Grigoriev}, where the notion of \textit{intrinsic} Lagrangian is introduced (see also~\cite{GriGri}). An \textit{intrinsic} Lagrangian originates from a single differential form representing an \textit{internal} Lagrangian. In this paper, we focus on variational principles originating from jets and explore how they affect the intrinsic geometry of equations. Therefore, we deal with equivalence classes and not with representatives that can be chosen ambiguously. The approach associated with \textit{intrinsic} Lagrangians can be said to deal with objects for each of which such a representative is fixed and considered part of its original structure. Such objects are canonically associated with bundles of a particular form and action functionals on their sections. The variation of an \textit{intrinsic} Lagrangian within the class of \textit{all} embeddings $N\to\mathcal{E}$ generates the variation of the canonically associated action. But we have to limit ourselves to almost Cartan embeddings since we study \textit{internal} Lagrangians (not representatives).\\[-1 ex] 

\remarka{One can define stationary points of conservation laws the same way (and then they are determined by the corresponding variational $1$-forms).}\\[-1 ex]

In the examples below, we use more suitable indices for local coordinates on jets and equations.\\[-1 ex]

\examplea{\label{ex1} Consider the infinite prolongation $\mathcal{E}$ of the Laplace equation 
$$
u_{xx} + u_{yy} = 0\,.
$$
Here $\pi\colon \mathbb{R}\times \mathbb{R}^2\to \mathbb{R}^2$ is the projection onto the second factor. Laplace's equation is the Euler-Lagrange equation for
$$
L = -\frac{u_x^2 + u_y^2}{2}\,dx\wedge dy\,.
$$
Formula~\eqref{Noeth} can be used to find an appropriate form $\omega_L$. Here we have
$$
\mathcal{L}_{E_{\varphi}}(L) = -\big(u_x D_x(\varphi) + u_y D_y(\varphi)\big)dx\wedge dy\,.
$$
Integrating by parts, we get
$$
\mathcal{L}_{E_{\varphi}}[L]_h = \langle \mathrm{E}[L]_h, \varphi \rangle + d_h[- u_x \varphi\, dy + u_y \varphi\, dx]_h\,.
$$
Then we can take $\omega_L = - u_x\,\theta_0\wedge dy + u_y\,\theta_0\wedge dx$, where $\theta_0 = du - u_x\, dx - u_y\, dy$. 
\noindent
The corresponding internal Lagrangian of the system $\mathcal{E}$ is represented by the restriction of the differential form
$$
L + \omega_L = -\frac{u_x^2 + u_y^2}{2}\,dx\wedge dy - u_x\,\theta_0\wedge dy + u_y\,\theta_0\wedge dx\,.
$$

One can regard the coordinate $u_{yy}$ and its derivatives as external coordinates for the corresponding infinite prolongation. Other coordinates on $J^{\infty}(\pi)$ can be treated as local coordinates on $\mathcal{E}$. Then the restrictions of the total derivatives to the system $\mathcal{E}$ have the form
\begin{align*}
&\,\overline{\!D}_x = \partial_x + u_x\partial_u + u_{xx}\partial_{u_x} + u_{xy}\partial_{u_y} + u_{xxx}\partial_{u_{xx}} + u_{xxy}\partial_{u_{xy}} + u_{xxxx}\partial_{u_{xxx}} + \ldots\,,\\
&\,\overline{\!D}_y = \partial_y + u_y\partial_u + u_{xy}\partial_{u_x} - u_{xx}\partial_{u_y} + u_{xxy}\partial_{u_{xx}} - u_{xxx}\partial_{u_{xy}} + u_{xxxy}\partial_{u_{xxx}} + \ldots
\end{align*}

One can take any compact $2$-dimensional submanifold $N\subset\mathbb{R}^2$. The specific choice of such a submanifold is not of great importance here. For definiteness, we choose the closed unit disk:
$$
N = \{(x, y)\in \mathbb{R}^2\, |\ x^2 + y^2\leqslant 1\}\,.
$$
Suppose $Y$ is a function on $N$ and $\sigma\in \mathcal{A}_N(\mathcal{E})$ is a (local) section of the bundle $\pi_{\infty}|_{\mathcal{E}}$ such that
\begin{align}
d\sigma(\partial_x + Y(x, y)\partial_y) \in \mathcal{C}_{\sigma(x, y)}\qquad \text{for all}\quad (x, y)\in N.
\label{covec1}
\end{align}
Then $d\sigma(\partial_x + Y\partial_y) = \,\overline{\!D}_x + Y\,\overline{\!D}_y$ for all $(x, y)\in N$, and $\sigma$ is of the form
\begin{align}
u = f\,,\qquad u_y = g\,, \qquad u_x = \partial_x f + Y(\partial_y f - g)\,,\qquad \ldots
\label{sigma}
\end{align}
Indeed, one way or another, $\sigma$ must determine some functions as $\sigma^*(u)$, $\sigma^*(u_y)$. The expressions for all other coordinates on $\mathcal{E}$ are uniquely defined due to~\eqref{covec1}. For instance, the relations
\begin{align*}
&(\,\overline{\!D}_x + Y\,\overline{\!D}_y)u_{x} = (\partial_x + Y\partial_y)\big(\partial_x f + Y(\partial_y f - g)\big)\,,\qquad (\,\overline{\!D}_x + Y\,\overline{\!D}_y)u_y = (\partial_x + Y\partial_y)g
\end{align*}
allow one to obtain the expressions for $u_{xx}$ and $u_{xy}$ since $\,\overline{\!D}_y(u_y) = -u_{xx}$ and the corresponding matrix of coefficients is non-degenerate. The same matrix appears when deriving expressions for higher order derivatives.
So, arbitrary smooth functions $f, g\colon N \to \mathbb{R}$ determine an appropriate $\sigma$ and vice versa. 

Denote by $\delta f$ and $\delta g$ arbitrary smooth functions on $N$ vanishing with all their derivatives on $\partial N$. Let us restrict our attention to paths in $\mathcal{A}_N(\mathcal{E})$ resulting from the replacing $f\mapsto f + \tau \delta f$, $g\mapsto g + \tau \delta g$ in~\eqref{sigma}:
\begin{align*}
u = f + \tau \delta f\,,\ \ u_y = g + \tau \delta g\,,\ \ u_x = \partial_x (f + \tau\delta f) + Y\big(\partial_y (f + \tau\delta f) - (g + \tau\delta g)\big)\,,\ \ \ldots
\end{align*}
Given this, it suffices to derive the Euler-Lagrange equations for the following formal expression
\begin{align*}
\gamma(0)^*(l) = \Big(\frac{Y^2 (\partial_y f - g)^2 - (\partial_x f)^2 + g^2}{2} - g\, \partial_y f\Big)\,dx\wedge dy\,,\qquad l = (L + \omega_L)|_{\mathcal{E}}\,.
\end{align*}
Varying w.r.t. the formal variables $f$ and $g$, we obtain the equations
\begin{align*}
\partial_x^2 f + \partial_y \kern 0.06em g + \partial_y\big(Y^2(g - \partial_y f)\big) = 0\,,\qquad (g - \partial_y f)(Y^2 + 1) = 0\,.
\end{align*}
Any functions $f$ and $g$ satisfying to these equations coincide with $s^*(u)$ and $s^*(u_y)$ for some solution $s\colon N \to \mathcal{E}$. Since they determine a unique appropriate local section, they give rise to the corresponding solution.

Therefore, an almost Cartan embedding $\sigma\in \mathcal{A}_N(\mathcal{E})$ that is a local section of the bundle $\pi_{\infty}|_{\mathcal{E}}$ and satisfies~\eqref{covec1} is a stationary point of the internal Lagrangian under consideration iff it defines a solution to the system $\mathcal{E}$. 
Furthermore, to obtain this result, it suffices to consider only paths that pass through local sections satisfying condition~\eqref{covec1} (i.e., in this example, we do not need to vary w.r.t. $Y$) and such that $\delta f$, $\delta g$ vanish with all their derivatives on $\partial N$.
As we will see below, this is not a coincidence (see Theorem~\ref{Nonchar}).
}\\[-1ex]

\section{\label{ILcl} Internal Lagrangians and conservation laws}

Conservation laws of an infinitely prolonged system of differential equations $\mathcal{E}$ are related to the trivial internal Lagrangian. More precisely, the differential 
$$
d_1\colon E^{\,0,\, n-1}_1(\mathcal{E}) \to E^{1,\, n-1}_1(\mathcal{E})
$$
maps conservation laws of $\mathcal{E}$ to variational $1$-forms. The group of variational $1$-forms $E^{1,\, n-1}_1(\mathcal{E})$ admits the following description
\begin{align*}
E^{1,\, n-1}_1(\mathcal{E}) = \dfrac{\{\omega\in\mathcal{C}\Lambda^n(\mathcal{E})\colon\ d\omega\in \mathcal{C}^2\Lambda^{n+1}(\mathcal{E})\}}
{\mathcal{C}^2\Lambda^{n}(\mathcal{E}) + d(\mathcal{C}\Lambda^{n-1}(\mathcal{E}))}\,.
\end{align*}
Therefore, each variational $1$-form defines an element of group~\eqref{SIL} and the corresponding internal Lagrangian. Any exact variational $1$-form $\vartheta\in \mathrm{im}\, d_1$ can be represented by an exact differential form; accordingly, the $\mathrm{im}\, d_1$ is contained in the kernel of the mapping $E^{1,\, n-1}_1(\mathcal{E}) \to \widetilde{E}^{\,0,\, n-1}_1(\mathcal{E})$.

It can be shown that a more general statement is true.\\[-1ex]

\propositiona{\label{ilclpro} A variational $1$-form $\vartheta\in E^{1,\, n-1}_1(\mathcal{E})$ produces the trivial internal Lagrangian if and only if $\vartheta\in \mathrm{im}\, d_1$.}\\[1ex]
\textbf{Proof.}{
The inclusions of the cochain complexes
$\Lambda^{*}(\mathcal{E}) \supset \mathcal{C}\Lambda^{*}(\mathcal{E}) \supset \mathcal{C}^2\Lambda^{*}(\mathcal{E})$
allow us to consider three short exact sequences
\begin{align*}
\xymatrix
{
0\ar[r] &\mathcal{C}\Lambda^{*}(\mathcal{E})\ar[r] &\Lambda^{*}(\mathcal{E}) \ar[r]
& \dfrac{\Lambda^{*}(\mathcal{E})}{\mathcal{C}\Lambda^{*}(\mathcal{E})} \ar[r] &0\,,
}\\
\xymatrix
{
0\ar[r] &\mathcal{C}^2\Lambda^{*}(\mathcal{E})\ar[r] &\Lambda^{*}(\mathcal{E}) \ar[r]
& \dfrac{\Lambda^{*}(\mathcal{E})}{\mathcal{C}^2\Lambda^{*}(\mathcal{E})} \ar[r] &0\,,
}\\
\xymatrix
{
0\ar[r] &\mathcal{C}^2\Lambda^{*}(\mathcal{E})\ar[r] &\mathcal{C}\Lambda^{*}(\mathcal{E}) \ar[r]
& \dfrac{\mathcal{C}\Lambda^{*}(\mathcal{E})}{\mathcal{C}^2\Lambda^{*}(\mathcal{E})} \ar[r] &0\,.
}
\end{align*}
Since the de Rham cohomology groups $H^{i}_{dR}(\mathcal{E})$ are trivial for $i > 0$ (according to Regularity assumptions, Section~\ref{difeq}), the corresponding long exact sequences have the form
\begin{align}
\xymatrix
{
\ldots \ar[r] & 0\ar[r] &E^{\,0,\, n-1}_1(\mathcal{E})\ar[r] & H^{n}(\mathcal{C}\Lambda^{*}(\mathcal{E})) \ar[r] &0 \ar[r]&\ldots\,,
}
\label{LES1}\\[1 ex]
\xymatrix
{
\ldots \ar[r] & 0\ar[r] &\widetilde{E}^{\,0,\, n-1}_1(\mathcal{E})\ar[r] & H^{n+1}(\mathcal{C}^2\Lambda^{*}(\mathcal{E})) \ar[r] &0 \ar[r]&\ldots\,,
}
\label{LES2}\\[1 ex]
\xymatrix
{
\ldots \ar[r] & H^{n}(\mathcal{C}\Lambda^{*}(\mathcal{E})) \ar[r] & E^{1,\, n-1}_1(\mathcal{E}) \ar[r] & H^{n+1}(\mathcal{C}^2\Lambda^{*}(\mathcal{E})) \ar[r] &\ldots
\nonumber
}
\end{align}
Here the mappings
$$
E^{\,0,\, n-1}_1(\mathcal{E})\to H^{n}(\mathcal{C}\Lambda^{*}(\mathcal{E}))\,,\qquad \widetilde{E}^{\,0,\, n-1}_1(\mathcal{E})\to H^{n+1}(\mathcal{C}^2\Lambda^{*}(\mathcal{E}))
$$
are induced by the de Rham differential $d$. Exact sequences~\eqref{LES1} and~\eqref{LES2} show that these mappings
are isomorphisms.
Thus, the assertion of the proposition follows from the long exact sequence (with the desired mappings)
\begin{align*}
\xymatrix
{
\ldots \ar[r] & E^{\,0,\, n-1}_1(\mathcal{E}) \ar[r]^{d_1} & E^{1,\, n-1}_1(\mathcal{E}) \ar[r] & \widetilde{E}^{\,0,\, n-1}_1(\mathcal{E}) \ar[r] &\ldots
}
\end{align*}
}\\[-5ex]

Hereby, a variational $1$-form is related to a conservation law iff it produces the trivial internal Lagrangian.

\section{\label{iLs} Internal Lagrangians and symmetries}

It is quite natural to expect that some version of Noether's theorem concerns internal Lagrangians. Let us recall that internal Lagrangians of a system of differential equations $\mathcal{E}$ are related to presymplectic structures via the mapping
$$
\tilde{d}^{\,0,\,n-1}_1\colon \widetilde{E}_1^{\,0,\,n-1}(\mathcal{E})\to E_1^{2,\,n-1}(\mathcal{E})
$$
induced by the de Rham differential $d$.

Symmetries of differential equations act on internal Lagrangians by means of the Lie derivative. Suppose $\boldsymbol\ell\in\widetilde{E}^{\,0,\,n-1}_1(\mathcal{E})$ is an internal Lagrangian, $l\in \boldsymbol\ell$ is a differential form representing $\boldsymbol\ell$, $X$ is a symmetry. The Lie derivative $\mathcal{L}_X(l)$ can be written as
$$
\mathcal{L}_X(l) = d(i_X l) + i_X dl\,.
$$
This means that the corresponding internal Lagrangian is represented by the differential form $i_X dl$. In other words, the internal Lagrangian $\mathcal{L}_X\, \boldsymbol\ell$ is produced by the variational $1$-form 
$$
i_X \tilde{d}^{\,0,\,n-1}_1 \boldsymbol\ell = i_X dl + \mathcal{C}^2\Lambda^{n}(\mathcal{E}) + d(\mathcal{C}\Lambda^{n-1}(\mathcal{E})).
$$
Then from Proposition~\ref{ilclpro} it follows that an internal Lagrangian $\boldsymbol\ell$ is invariant under the action of a symmetry $X$ (i.e., $\mathcal{L}_X\, \boldsymbol\ell = 0$) iff $X$ produces conservation laws. We get the following version of Noether's theorem\\[-1ex]

\theorema{(Noether's theorem). Let $\boldsymbol\ell\in \widetilde{E}_1^{\,0,\,n-1}(\mathcal{E})$ be an internal Lagrangian, and let $X$ be a symmetry of an infinitely prolonged system of differential equations $\mathcal{E}$. If $\boldsymbol\ell$ is invariant under the action of $X$, then the variational $1$-form $\vartheta = i_{X}\, \tilde{d}^{\, 0,\,n-1}_1 \boldsymbol\ell$ is exact, and $X$ gives rise to the conservation laws $d_1^{\,-1}(\vartheta)$. Otherwise, $X$ produces the non-trivial internal Lagrangian $\mathcal{L}_X\, \boldsymbol\ell$.}\\[-1ex]

If an infinitely prolonged system of equations $\mathcal{E}\subset J^{\infty}(\pi)$ admits gauge symmetries and its projection to $J^0(\pi)$ is surjective, then \textit{all internal Lagrangians of $\mathcal{E}$ are gauge invariant}, as are all of its presymplectic structures. This follows from the fact that any gauge symmetry produces the trivial variational $1$-form. Indeed, 
assume that $\Delta$ is an operator representing a presymplectic structure, $R$ is an operator that generates gauge symmetries of $\mathcal{E}$. A gauge symmetry $R(\epsilon)$ gives rise to the cosymmetry $\psi = \Delta (R\,(\epsilon))$.
According to Green's formula, there exists a $\mathcal{C}$-differential operator $\square\colon P(\mathcal{E})\to \Lambda^{n-1}_h(\mathcal{E})$ such that
$$
\langle \psi, G \rangle = \langle \epsilon, R^*\circ \Delta^*\, (G) \rangle  + d_h\,\square(G)\qquad \text{for all}\quad G\in P(\mathcal{E})\,.
$$
Since $R^*\circ \Delta^*\circ l_{\mathcal{E}} = R^*\circ l^{\,*}_{\mathcal{E}} \circ \Delta = (l_{\mathcal{E}}\circ R)^* \circ \Delta = 0$,
we can use the operator $\square\circ l_{\mathcal{E}}$ as $\nabla$ in~\eqref{cos1f}. However, operators of the form 
$$
A\circ l_{\mathcal{E}}\,,\qquad A\in \mathcal{C}(P(\mathcal{E}), \Lambda^{n-1}_h(\mathcal{E}))
$$
determine the trivial element of $E^{1,\, n-1}_0(\mathcal{E})$.\\[-1ex]

\remarka{One can also apply the $k$-line theorem (see, e.g.,~\cite{KraVer2}) to show that any gauge symmetry produces the trivial variational $1$-form.}\\

\section{\label{Compl} Internal Lagrangians and non-degenerate forms}

Let $\pi\colon E\to M$ be a locally trivial smooth vector bundle over a smooth manifold $M$, $\dim M = n$, $\dim E = n + m$. Suppose $N\subseteq M$ is a compact oriented $n$-dimensional submanifold. If $\mathcal{E}\subset J^{\infty}(\pi)$ is an infinitely prolonged system of differential equations, then it can be endowed with the bundle structure
$$
\pi_{\mathcal{E}}\colon \mathcal{E}\to M,\qquad \pi_{\mathcal{E}} = \pi_{\infty}|_{\mathcal{E}}.
$$

\definitiona{Let $\sigma\colon N\to \mathcal{E}$ be an almost Cartan embedding. We say that $\sigma$ is an \textit{almost Cartan section} of $\pi_{\mathcal{E}}$ if $\pi_{\mathcal{E}}\circ \sigma = \mathrm{id}_N$.}\\[-1ex]

\remarka{In general, point transformations of jets do not preserve the bundle structure (as well as contact ones). Therefore, they induce transformations of almost Cartan sections to almost Cartan embeddings, which are not necessarily almost Cartan sections.}\\[-1ex] 

\noindent
Almost Cartan sections are also related to initial-boundary conditions. It is reasonable to anticipate that characteristic planes somehow manifest themselves with variations of internal Lagrangians.\\[-1 ex]

\definitiona{Let $k\geqslant 1$ be an integer. A differential $n$-form $L\in \Lambda^n(J^{k}(\pi))$ is \textit{non-degenerate} if, for each point of the system of equations $\{\mathrm{E}[L]_h = 0\}\subset J^{2k}(\pi)$, there exists a non-characteristic (and non-zero) covector.
}\\[-1ex]

We shall say that the Euler-Lagrange equations $\mathrm{E}[L]_h = 0$ for a non-degenerate differential form $L$ are also non-degenerate. If $L\in \Lambda^n(J^{k}(\pi))$ is of the form $\lambda\, dx^1\wedge\ldots \wedge dx^n$, then the non-degeneracy condition is equivalent to the following one in adapted coordinates. For each $p\in \{\mathrm{E}[L]_h = 0\}$, there exists a covector $\xi = \xi_i\, dx^i\in T^{\,*}_{\pi_{2k}(p)} M$ such that the matrix
\begin{align*}
A_{ij} = \sum_{|\alpha| = |\beta| = k}\, \dfrac{\partial^2 \lambda}{\partial u^i_{\alpha} \partial u^j_{\beta}}\,\xi^{\alpha + \beta}
\end{align*}
has a non-zero determinant at $p$.
Here $\beta$ denotes a multi-index: $\beta = \beta_1 x^1 + \ldots + \beta_n x^n$,
$$
\xi^{\alpha + \beta} = (\xi_1)^{\alpha_1 + \beta_1}\cdot\ldots\cdot (\xi_n)^{\alpha_n + \beta_n}\,.
$$

\definitiona{Let $\xi\in\Lambda^1(M)$ be a covector field. An almost Cartan section $\sigma\colon N\to \mathcal{E}$ is a $\xi$-\textit{section} of $\pi_{\mathcal{E}}$ if
$d\sigma(\ker \xi|_x) \subset \mathcal{C}_{\sigma(x)}$ for all $x\in N$.\\[-1ex]
}


Non-degenerate forms play an important role in the covariant phase space formalism. Appropriate integral surfaces of an involutive distribution of the form $\xi = 0$ can be viewed as instantaneous surfaces for some space+time decomposition of $M$. Let us note that, in the case of stationary points, perturbations of spatial distributions are also allowed. In this context, the cases $n = 1, 2$ stand out from the rest since, for $n > 2$, not all codimension-$1$ distributions are involutive. 
\\[-1ex]

\definitiona{Let $\boldsymbol\ell$ be an internal Lagrangian of an infinitely prolonged system of equations $\mathcal{E}$, $l\in \boldsymbol\ell$ be a differential form representing $\boldsymbol\ell$. We say that a $\xi$-section $\sigma$ is a $\xi$-\textit{stationary point} of $\boldsymbol\ell$ if
\begin{align*}
\dfrac{d}{d\tau}\Big|_{\tau = 0}\int_N \gamma(\tau)^*(l) = 0
\end{align*}
for any path $\gamma\colon \mathbb{R}\times N\to \mathcal{E}$ in $\xi$-sections of $\pi_{\mathcal{E}}$ such that $\gamma(0) = \sigma$ and each point of the boundary is fixed.
}\\[-1ex]


\remarka{If $l$ is a differential form representing an internal Lagrangian of an infinitely prolonged system $\mathcal{E}$, $\gamma$ is a path in $\mathcal{A}_N(\mathcal{E})$, and $h(\tau)$ is a family of orientation-preserving automorphisms of $N$ (for $\tau\in\mathbb{R}$) such that each point of the boundary is fixed, then
\begin{align*}
\int_N \gamma(\tau)^*(l) = \int_N \big(\gamma(\tau)\circ h(\tau)\big)^*(l)\qquad \text{for each}\quad \tau\in\mathbb{R}\,.
\end{align*}
So, speaking about stationary points of internal Lagrangians, we do not need to consider paths in $\mathcal{A}_N(\mathcal{E})$ obtained from ones in $\xi$-sections by compositions with such isotopies.}\\[-1ex]

If a $\xi$-section is a stationary point of an internal Lagrangian, it is also a $\xi$-stationary one. We propose a theorem establishing a connection between non-degenerate forms and the corresponding internal Lagrangians in terms of $\xi$-stationary points. For convenience, we have divided the proof into seven parts.\\[-1ex]

\theorema{\label{Nonchar} Let $L\in \Lambda^n(J^{k}(\pi))$ be a non-degenerate form, and let $\mathcal{E}$ be the infinite prolongation of the system of Euler-Lagrange equations $\mathrm{E}[L]_h = 0$. Suppose $\xi\in\Lambda^1(M)$ is a non-vanishing covector field such that the distribution $\xi = 0$ is involutive, and for each $p\in \{\mathrm{E}[L]_h = 0\}$, the covector $\xi|_{\pi_{2k}(p)}$ is non-characteristic. Then a $\xi$-section $\sigma$ is a $\xi$-stationary point of the corresponding internal Lagrangian if and only if $\sigma$ is a (local) solution to $\pi_{\mathcal{E}}$.}\\[-2ex]

\noindent
\textbf{Proof. }{According to Proposition~\ref{Sol}, any $\xi$-section that defines a solution to $\pi_{\mathcal{E}}$ is a $\xi$-stationary point. It suffices to prove that each $\xi$-stationary point defines a solution in a neighborhood of any interior point of its domain.\\[0.5ex]
\textit{\textbf{1.}} Let $N\subseteq M$ be an $n$-dimensional compact submanifold. Fix an interior point $x_0\in \mathrm{int}\, N$. Since the distribution $\xi = 0$ is involutive, we can choose local coordinates $x^1$, \ldots, $x^n$ in a neighborhood $U\subset N$ of $x_0$ such that the distributions $\xi = 0$ and  
$dx^n = 0$ coincide. We shall use the notation $x^n = y$ and the \textit{``spatial'' multi-index $s = s_1 x^1 + \ldots + s_{n-1} x^{n-1}$}.

Without loss of generality, we can assume that $L$ is of the form $L = \lambda\, dx^1\wedge\ldots \wedge dx^n$.
The covector $dy$ is non-characteristic at every point of the system $\{\mathrm{E}[L]_h = 0\}$ over $U$. This implies that the determinant of the matrix
$$
A_{ij} = \dfrac{\partial^2 \lambda}{\partial u^i_{ky} \partial u^j_{ky}}
$$
never vanishes in $\pi_{2k}^{-1}(U)$. Then, in $\pi_{2k}^{-1}(U)$, the quasilinear system $\mathrm{E}[L]_h = 0$ is equivalent to a system of the form
$u^i_{2ky} = h^i$.
The functions $h^i$ do not depend on $u^1_{2ky}$, \ldots, $u^m_{2ky}$ and their derivatives. Hence, $x^1$, \ldots, $x^n$, and $u^i_{s + jy}$ ($i = 1, \ldots, m$; $j = 0, \ldots, 2k-1$) can be regarded as coordinates in $\pi_{\infty}^{-1}(U)\cap\mathcal{E}$.
Denote by $\,\overline{\!D}_{x^1}$, $\ldots$\,, $\,\overline{\!D}_{x^n}$ the corresponding total derivatives on $\pi_{\infty}^{-1}(U)\cap\mathcal{E}$.
From now on, we use only these local coordinates.\\[0.5ex]
\textit{\textbf{2.}} In the local coordinates, any $\xi$-section $N\to \mathcal{E}$ has the form $u^i_{s + jy} = \partial_s f^i_j$
for some functions $f^i_j\colon U\to \mathbb{R}$. Such functions define a section of the trivial bundle
\begin{align*}
\mathrm{pr}_U \colon \mathbb{R}^{2km}\times U\to U\,.
\end{align*}
Denote by $v^i_j$ coordinates along the fibers of $\mathrm{pr}_U$ and define the fiberwise mapping $\Phi\colon J^{\infty}(\mathrm{pr}_U)\to \mathcal{E}$ by $u^i_{s + jy} = v^i_{j;\, s}$\,, where $v^i_{j;\,\alpha}$ are adapted coordinates on $J^{\infty}(\mathrm{pr}_U)$. We shall use the notation $d/dx^1$, $\ldots$\,, $d/dx^n$ for the corresponding total derivatives on~$J^{\infty}(\mathrm{pr}_U)$.

Let $\sigma\colon N\to \mathcal{E}$ be a $\xi$-stationary point. There exists a unique section $f_{\sigma}$ of the bundle $\mathrm{pr}_U$ such that $\sigma|_{U} = \Phi\circ j^{\infty}\!f_{\sigma}$. Fix a compact submanifold $N'\subset U$ such that $x_0\in \mathrm{int}\, N'$ and consider a smooth path $\widetilde{\gamma}$ in sections of $\mathrm{pr}_U$. Suppose $\widetilde{\gamma}$ has the form
\begin{align*}
\widetilde{\gamma}\colon \qquad v^i_j = {f_{\sigma}}^i_j + \tau\delta f^i_j\,,
\end{align*}
where $\delta f^i_j$ denote arbitrary smooth functions on $U$ that vanish with all their derivatives on $\partial N'$ and are equal to zero in $U\setminus N'$. For every $\tau\in \mathbb{R}$, we can extend $\Phi\circ j^{\infty} \widetilde{\gamma}(\tau, \cdot )$ to $N$ so that the extension coincides with $\sigma$ in  $N\setminus U$. The corresponding path in $\xi$-sections is smooth. Therefore, in $N'$, $f_{\sigma}$ must satisfy the Euler-Lagrange equations for $[\Phi^*(l)]_h$ since $\sigma$ is a $\xi$-stationary point.

So, it suffices to show that each solution to the infinitely prolonged Euler-Lagrange equations for the horizontal form $[\Phi^*(l)]_h$ defines a local solution to $\pi_{\mathcal{E}}$.\\[0.5ex]
\textit{\textbf{3.}} We seek a Cartan $n$-form $\omega_L\in\mathcal{C}\Lambda^n(\pi)$ satisfying identity~\eqref{Noeth}. An appropriate form can be found using integration by parts. The component of the Lie derivative $\mathcal{L}_{E_{\varphi}}(L)$ reads
\begin{align*}
\dfrac{\partial \lambda}{\partial u^i_{\alpha}}\, D_{\alpha}(\varphi^i) = &\ D_{x^1}(\ldots) + \ldots + D_{x^{n-1}}(\ldots) + (-1)^{|s|}D_{s}
\Big(\dfrac{\partial \lambda}{\partial u^i_{s + jy}}\Big)D_{jy}(\varphi^i) + \textit{linear in }\varphi.
\end{align*}
Here we have applied integration by parts w.r.t. the variables $x^1$, \ldots, $x^{n-1}$. All the summands linear in $\varphi$ will end up in the expression for the variational derivative, so they are of no interest to us. Besides, $d\Phi(d/dx^i) = \,\overline{\!D}_{x^i}$ for $i = 1, \ldots, n-1$. Therefore, if a Cartan $n$-form $\chi\in\mathcal{C}\Lambda^n(\mathcal{E})$ is proportional to $dy$, then $[\Phi^*(\chi)]_h = 0$.
This observation explains why we are interested only in summands of the form~$D_y(\ldots)$.\\[0.5ex]
\textit{\textbf{4.}} Completing the integration by parts, we obtain:
\begin{align*}
\begin{aligned}
\dfrac{\partial \lambda}{\partial u^i_{\alpha}}\, D_{\alpha}(\varphi^i) =&\ (-1)^{|\alpha|}D_{\alpha}\Big(\dfrac{\partial \lambda}{\partial u^i_{\alpha}}\Big)\, \varphi^i + D_y(B) +{}\\
&+ D_{x^1}(\ldots) + \ldots + D_{x^{n-1}}(\ldots)\,,
\end{aligned}
\qquad\quad
B = (-1)^{|\alpha|}D_{\alpha}\Big(\dfrac{\partial \lambda}{\partial u^i_{\alpha + (r+1)y}}\Big)D_{ry}(\varphi^i).
\end{align*}
So, the corresponding internal Lagrangian is represented by the form $l = (L + \omega_L)|_{\mathcal{E}}$, where
\begin{align*}
\omega_L =&\ dx^1\wedge\ldots\wedge dx^{n-1}\wedge (-1)^{|\alpha|}D_{\alpha}\Big(\dfrac{\partial \lambda}{\partial u^i_{\alpha + (r+1)y}}\Big)\,\theta^i_{ry} +\ \textit{proportional to}\ dy.
\end{align*}
It is easy to see that the maximum value of the summation index $r$ in the last formula is equal to $k-1$ (while the minimum value is $0$). Then the differential form
\begin{align}
\Big(\Phi^*(\lambda) + (-1)^{|\alpha|}\widetilde{D}_{\alpha}\Big(\dfrac{\partial \lambda}{\partial u^i_{\alpha + (r+1)y}}\Big)(v^i_{r;\, y} - v^i_{r+1})\Big)dx^1\wedge\ldots\wedge dx^n\,,\qquad \widetilde{D}_{\alpha} = \Phi^*\circ D_{\alpha}\,.
\label{pullback}
\end{align}
represents the horizontal form $[\Phi^*(l)]_h$.
Now it only remains to derive the Euler-Lagrange equations for the $[\Phi^*(l)]_h$.\\[0.5ex]
\textit{\textbf{5.}} Below we deal with the component of differential form~\eqref{pullback} and the action it determines.

Consider the case $k > 2$. The variables $v^j_{2k-1}$ appear only in the terms corresponding to $\alpha = (k-1)y$, $r = 0$ (their derivatives do not appear in~\eqref{pullback} at all). The variational derivatives $\delta/\delta v^j_{2k-1}$ yield the equations
\begin{align*}
(-1)^{k-1}\,\Phi^*\Big(\dfrac{\partial^2 \lambda}{\partial u^j_{ky} \partial u^i_{ky}}\Big)
(v^i_{0;\, y} - v^i_{1}) = 0.
\end{align*}
The non-degeneracy condition implies $v^i_{1} = v^i_{0;\, y}$.

Let us recall that the formula for the variational derivative with respect to a variable is a sum of compositions of derivatives (partial and total ones). From now on, in the terms corresponding to $r = 0$, we apply partial derivatives only to the factors $v^i_{0;\, y} - v^i_{1}$; otherwise we would get trivial summands because of the equations $v^i_{1} = v^i_{0;\, y}$ and their differential consequences.

Among the terms corresponding to $r > 0$, the variables $v^j_{2k-2}$ appear only in the summands corresponding to $\alpha = (k-2)y$, $r = 1$ (their derivatives do not appear whenever $r > 0$). Thus, varying w.r.t. $v^j_{2k-2}$, we get
\begin{align*}
(-1)^{k-2}\,\Phi^*\Big(\dfrac{\partial^2 \lambda}{\partial u^j_{ky} \partial u^i_{ky}}\Big)
(v^i_{1;\, y} - v^i_{2}) = 0.
\end{align*}
The non-degeneracy condition implies $v^i_{2} = v^i_{1;\, y}$. From now on, in the terms corresponding to $r = 1$, we apply partial derivatives only to the factors $v^i_{1;\, y} - v^i_{2}$.

Following the same line of reasoning for $\delta/\delta v^j_{2k-3}$, $\ldots\,$, $\delta/\delta v^j_{k+1}$, we obtain the equations
\begin{align*}
v^i_{1} = v^i_{0;\, y}\,,\qquad v^i_{2} = v^i_{1;\, y}\,, \qquad\ldots\,,\qquad v^i_{k-1} = v^i_{k-2;\, y}\,.
\end{align*}
\textit{\textbf{6.}} Let us discuss the variational derivative $\delta/\delta v^j_{k}$ (still for~\eqref{pullback}). It suffices to consider the term $\Phi^*(\lambda)$ and the terms corresponding to $r = k-1$, $\alpha = 0$. Then we find
\begin{align*}
\Phi^*\Big(\dfrac{\partial^2 \lambda}{\partial u^j_{ky} \partial u^i_{ky}}\Big)
(v^i_{k-1;\, y} - v^i_{k}) = 0\,,
\end{align*}
the other terms cancelling. These equations imply that $v^i_{k} = v^i_{k-1;\, y}$. From now on, in the sum
\begin{align*}
(-1)^{|\alpha|}\widetilde{D}_{\alpha}\Big(\dfrac{\partial \lambda}{\partial u^i_{\alpha + (r+1)y}}\Big)(v^i_{r;\, y} - v^i_{r+1})\,,
\end{align*}
we apply partial derivatives only to the factors of the form $v^i_{r;\, y} - v^i_{r+1}$.

We can consider only the terms corresponding to $r = k-1,\, k-2$ and the $\Phi^*(\lambda)$ when varying w.r.t. $v^j_{k-1}$.
Note that $\Phi^*(\lambda)$ does not depend on coordinates of the form $v^j_{r;\, \alpha + y}$.
Then we~get
\begin{align*}
0 &= (-1)^{|s|}\widetilde{D}_{s}\Big(\dfrac{\partial \lambda}{\partial u^j_{s + (k-1)y}}\Big) - (-1)^{|\alpha|}\widetilde{D}_{\alpha}\Big(\dfrac{\partial \lambda}{\partial u^j_{\alpha + (k-1)y}}\Big) - (-1)^{|\alpha|} \dfrac{d}{dy}\, \widetilde{D}_{\alpha}\Big(\dfrac{\partial \lambda}{\partial u^j_{\alpha + ky}}\Big) ={}\\
&= (-1)^{|\beta|}\widetilde{D}_{\beta + y}\Big(\dfrac{\partial \lambda}{\partial u^j_{\beta + ky}}\Big) - (-1)^{|\alpha|} \dfrac{d}{dy}\, \widetilde{D}_{\alpha}\Big(\dfrac{\partial \lambda}{\partial u^j_{\alpha + ky}}\Big)
= \widetilde{D}_{y}\Big(\dfrac{\partial \lambda}{\partial u^j_{ky}}\Big) - \dfrac{d}{dy}\, \Phi^*\Big(\dfrac{\partial \lambda}{\partial u^j_{ky}}\Big).
\end{align*}
However, we have already deduced the equations $v^i_{r;\, \alpha} = v^i_{0;\, \alpha + ry}$ for $r\leqslant k$. For this reason, the only non-trivial summands here are
\begin{align*}
\Phi^*\Big(\dfrac{\partial^2 \lambda}{\partial u^i_{ky} \partial u^j_{ky}}\Big)
(v^i_{k+1} - v^i_{k;\, y}).
\end{align*}
Hence, the equations $v^i_{r;\, \alpha} = v^i_{0;\, \alpha + ry}$ hold for all $r\leqslant k + 1$.

Again, we can consider only the terms corresponding to $r = k-2,\, k-3$ and the $\Phi^*(\lambda)$ when varying w.r.t. $v^j_{k-2}$. Similarly we obtain
\begin{align*}
0 &= (-1)^{|s|}\widetilde{D}_{s}\Big(\dfrac{\partial \lambda}{\partial u^j_{s + (k-2)y}}\Big) - (-1)^{|\alpha|}\widetilde{D}_{\alpha}\Big(\dfrac{\partial \lambda}{\partial u^j_{\alpha + (k-2)y}}\Big) - (-1)^{|\alpha|}\dfrac{d}{dy}\,\widetilde{D}_{\alpha}\Big(\dfrac{\partial \lambda}{\partial u^j_{\alpha + (k-1)y}}\Big) =\\
&= (-1)^{|\beta|}\widetilde{D}_{\beta + y}\Big(\dfrac{\partial \lambda}{\partial u^j_{\beta + (k-1)y}}\Big) - (-1)^{|\alpha|}\dfrac{d}{dy}\,\widetilde{D}_{\alpha}\Big(\dfrac{\partial \lambda}{\partial u^j_{\alpha + (k-1)y}}\Big).
\end{align*}
In this formula, one can express in $v^i_{0;\, \beta}$ everything except $\Phi^*(u^i_{(k+2)y})$ using the equations
\begin{align}
v^i_{r;\, \alpha} = v^i_{0;\, \alpha + ry}\qquad\quad  r\leqslant k + 1.
\label{relat}
\end{align} 
Upon substituting, we get
\begin{align*}
- \Phi^*\Big(\dfrac{\partial^2 \lambda}{\partial u^i_{ky} \partial u^j_{ky}}\Big)\bigg|_{\eqref{relat}}\!
\cdot (v^i_{k+2} - v^i_{0;\, (k+2)y}) = 0.
\end{align*}
So, the equations $v^i_{r;\, \alpha} = v^i_{0;\, \alpha + ry}$ hold for all $r\leqslant k + 2$.

Similar reasoning gives the equations $v^i_{k+3;\, \alpha} = v^i_{0;\, \alpha + (k+3)y}$\,, $\ldots$\, Finally, we obtain
\begin{align}
v^i_{r;\, \alpha} = v^i_{0;\, \alpha + ry}\qquad \text{for all} \quad r\leqslant 2k-1.
\label{complrelat}
\end{align}
\textit{\textbf{7.}} Now we know that all $\xi$-stationary points of the internal Lagrangian 
$$
\boldsymbol\ell = l + \mathcal{C}^2\Lambda^{n}(\mathcal{E}) + d(\Lambda^{n-1}(\mathcal{E}))
$$
define sections of $\mathrm{pr}_U$ that satisfy to equations~\eqref{complrelat}. Informally, we can say that $\xi$-stationary points satisfy equations~\eqref{complrelat}. The cases $k = 1$ and $k = 2$ do not differ from those considered above. In these cases, one can derive the same result by direct calculation.

Let us discuss the variational derivative $\delta/\delta v^j_0$. Substituting $v^i_{0;\, \alpha + ry}$ for $v^i_{r;\, \alpha}$ in~\eqref{pullback}, we obtain the original Lagrangian written in terms of $v^i_0$ instead $u^i$. This means: when we consider variations of $\xi$-sections satisfying equations~\eqref{complrelat} within the class of $\xi$-sections that also satisfy~\eqref{complrelat}, they
lead to the original Euler-Lagrange equations $\mathrm{E}[L]_h = 0$ expressed in $v^i_0$. Thus, any $\xi$-stationary point of the internal Lagrangian 
$\boldsymbol\ell$
must be a (local) solution of $\pi_{\mathcal{E}}$. 
This result completes the proof.}\\[-1ex]

All non-degenerate Euler-Lagrange equations are $l$-normal~\cite{VinKr}. Indeed, at least locally, such equations can be written in an extended (= generalized) Kovalevskaya form. This implies that such equations do not admit gauge symmetries. Prominent examples of degenerate non-gauge Euler-Lagrange equations are the Proca theory (see Section~\ref{degen}) and the massive spin-$2$ theory~\cite{FiePau}.\\[-1ex]

\remarka{In the case of gauge systems, one can consider perturbations of almost Cartan embeddings satisfying a gauge fixing condition within the class of all almost Cartan embeddings.}

\section{\label{Exam} Examples}

Example~\ref{ex1} from Section~\ref{VariL} illustrates Theorem~\ref{Nonchar}. We now consider several simple examples where the conditions of Theorem~\ref{Nonchar} are not satisfied.\\[-1ex]

\examplea{Let us see how characteristics of differential equations can interfere with ``completeness'' of internal Lagrangians.
Consider the wave equation
\begin{align*}
u_{xy} = 0
\end{align*}
and its infinite prolongation $\mathcal{E}$. Here $\pi\colon \mathbb{R}\times \mathbb{R}^2\to \mathbb{R}^2$ is the projection onto the second factor. 

This equation is the Euler-Lagrange equation for the non-degenerate $2$-form
\begin{align*}
L = -\dfrac{u_x u_y}{2}\, dx\wedge dy\,.
\end{align*} 
The variables $x$, $y$, $u$, $u_x$, $u_y$ $u_{xx}$, $u_{yy},\, \ldots$ can be viewed as local coordinates on $\mathcal{E}$. The restrictions of the total derivatives to the system $\mathcal{E}$ have the form
\begin{align*}
&\,\overline{\!D}_x = \partial_x + u_x\partial_u + u_{xx}\partial_{u_x} + u_{xxx}\partial_{u_{xx}} + \ldots\,,\qquad
\,\overline{\!D}_y = \partial_y + u_y\partial_u + u_{yy}\partial_{u_y} + u_{yyy}\partial_{u_{yy}} + \ldots
\end{align*}

The covector field $dy$ is characteristic for the wave equation. The total derivative $\,\overline{\!D}_x$ is tangent to the image of any $dy$-section of $\pi_{\mathcal{E}}$. Suppose $N\subset \mathbb{R}^2$ is a compact $2$-dimensional submanifold. A $dy$-section $\sigma\colon N\to\mathcal{E}$ has the form
\begin{align*}
u = f,\ \ u_x = \partial_x f,\ \ u_y = h_1, \ \ u_{xx} = \partial_x^{\,2} f, \ \ u_{yy} = h_2, \ \ u_{xxx} = \partial_x^{\,3} f, \ \ u_{yyy} = h_3,\ \ \ldots,
\end{align*}
Here $f$, $h_1$, $h_2,\, \ldots$ are functions on $N$. The function $f$ can be chosen arbitrarily, while the functions $h_1$, $h_2,\, \ldots$ do not depend  on the variable $x$.

The corresponding internal Lagrangian is represented by the form
$$
l = -\frac{u_x u_y}{2}\,dx\wedge dy - \dfrac{u_y}{2}\,\theta_0\wedge dy - \dfrac{u_x}{2}\,dx\wedge \theta_0\,.
$$
Let $\gamma\colon \mathbb{R}\times N\to \mathcal{E}$ be a path in $dy$-sections. In view of the fixed boundary points requirement, we can assume without loss of generality that $\gamma$ has the form
\begin{align*}
u = f + \tau \delta f,\ \ u_x = \partial_x (f + \tau \delta f),\ \ u_y = h_1, \ \ u_{xx} = \partial_x^{\,2} (f + \tau \delta f), \ \ u_{yy} = h_2,\ \ \ldots\,,
\end{align*}
where $\delta f$ is an arbitrary function on $N$ vanishing with all its derivatives w.r.t. $x$ on $\partial N$.
Then the pullback $\gamma(\tau)^*(l)$ reads
\begin{align*}
& - \dfrac{\partial_x (f + \tau\delta f)\, \partial_y (f + \tau\delta f)}{2}\, dx\wedge dy\,,
\end{align*}
and we obtain only one equation: $\partial_y\,\partial_x f = 0$.

We haven't got any equations for the functions $h_i$. Strictly speaking, an infinite number of relations is required here. It is clear that using an internal Lagrangian, it is impossible to obtain an infinite number of equations. This means that the set of all $dy$-stationary points of the internal Lagrangian contains more than just local solutions of $\pi_{\mathcal{E}}$.}\\[-1ex]

\examplea{There are examples of other manifestations of characteristics.
Suppose $\pi\colon \mathbb{R}^2\times \mathbb{R}^2 \to \mathbb{R}^2$ is the projection onto the second factor. Consider the non-degenerate differential $2$-form 
\begin{align}
L = \Big(\dfrac{u_t v - u v_t}{2} - \dfrac{u_x^2 + v_x^2}{2} - V(x)\,\dfrac{u^2 + v^2}{2}\Big)dt\wedge dx\,,
\label{Schro}
\end{align}
where $V(x)$ is an arbitrary function. This differential form leads to the $1$-dimensional Schr\"{o}dinger equation for $\hbar = 1$, $m = 1/2$, $\Psi = u + iv$\,:
\begin{align*}
&- v_t + u_{xx} - V(x)u = 0\,,\\
&u_t + v_{xx} - V(x)v = 0\,.
\end{align*}

The Schr\"{o}dinger equation has a unique characteristic distribution determined by the covector field $dt$. Suppose $N\subset \mathbb{R}^2$ is a compact $2$-dimensional submanifold. Choosing $t$, $x$, $u$, $v$ and the derivatives with respect to $x$ as local coordinates on the infinite prolongation $\mathcal{E}$, we see that any $dt$-section $\sigma\colon N\to \mathcal{E}$ has the form
\begin{align*}
u = f,\qquad v = g, \qquad u_x = \partial_x f,\qquad v_x = \partial_x \kern 0.05em g,\qquad u_{xx} = \partial_x^2 f,\qquad v_{xx} = \partial_x^2 \kern 0.05em g,\qquad \ldots
\end{align*}
The functions $f$ and $g$ are arbitrary functions on $N$.

The desired internal Lagrangian is represented by the form
\begin{align*}
l = - \dfrac{u_x^2 + uu_{xx} + v_x^2 + vv_{xx}}{2}\, dt\wedge dx + \dfrac{1}{2}(v\,\theta^u_0 - u\,\theta^v_0)\wedge dx - dt\wedge(u_x\,\theta^u_0 + v_x\,\theta^v_0)\,,
\end{align*}
where $\theta^u_0 = du - u_x\, dx + (v_{xx} - V(x)v)dt$ and $\theta^v_0 = dv - v_x\, dx - (u_{xx} - V(x)u)dt$.

Assume that $\gamma\colon \mathbb{R}\times N\to \mathcal{E}$ is a path in $dt$-sections such that each point of the boundary is fixed. It suffices to consider $\gamma$ of the form
\begin{align*}
u = f + \tau \delta f,\quad v = g + \tau \delta g,\quad u_x = \partial_x (f + \tau \delta f),\quad v_x = \partial_x(g + \tau \delta g),\quad  \ldots
\end{align*}
Here $\delta f$ and $\delta g$ are arbitrary functions on $N$ vanishing with all their derivatives w.r.t. $x$ on $\partial N$.
We~find
\begin{align*}
\gamma(0)^*(l) = \Big(\dfrac{g \partial_t f - f \partial_t g}{2} - \dfrac{(\partial_x f)^2 + (\partial_x g)^2}{2} - V(x)\,\dfrac{f^2 + g^2}{2}\Big) dt\wedge dx.
\end{align*}
Comparing this pullback with~\eqref{Schro}, we can conclude that all $dt$-stationary points of the internal Lagrangian are local solutions to $\pi_{\mathcal{E}}$.
}\\[-1ex]

\remarka{One can identify $dt$-sections of an evolutionary system of equations with sections of its original bundle $\pi\colon E\to M$. Then the variation of a Lagrangian of such a system has the same meaning as the variation of the corresponding internal Lagrangian within the class of $dt$-sections.}\\[-1ex]

\examplea{Let us go back to Example~\ref{ex1}. The Laplace equation admits the conservation law represented by the differential form
\begin{align*}
- u_y\, dx + u_x\, dy\,.
\end{align*}
This conservation law determines the differential covering~\cite{VinKr, KraVer1}
\begin{align*}
u_{xx} + u_{yy} = 0\,,\qquad v_x = -u_y\,,\qquad v_y = u_x\,.
\end{align*}
So, we can lift the internal Lagrangian form Example~\ref{ex1} to the Cauchy-Riemann equations
$$
u_y = -v_x\,,\qquad v_y = u_x\,.
$$ 

The variables $x$, $y$, $u$, $v$ and the derivatives with respect to $x$ can be treated as
local coordinates on the infinite prolongation $\mathcal{S}$ of the Cauchy-Riemann equations. The lift of the internal Lagrangian form Example~\ref{ex1} is generated by the form
\begin{align*}
\tilde{l} = -\frac{u_x^2 + v_x^2}{2}\,dx\wedge dy - u_x\,\theta^u_0\wedge dy - v_x\,\theta^u_0\wedge dx.
\end{align*}
Here $\theta_0^u = du - u_x\, dx + v_x\, dy$.
The restrictions of the total derivatives to $\mathcal{S}$ are
\begin{align*}
&\,\overline{\!D}_x = \partial_x + u_x\partial_u + v_x\partial_v + u_{xx}\partial_{u_x} + v_{xx}\partial_{v_x} + \ldots\,,\quad
\,\overline{\!D}_y = \partial_y - v_x\partial_u + u_x\partial_v - v_{xx}\partial_{u_x} + u_{xx}\partial_{v_x} + \ldots
\end{align*}

Let $N\subset \mathbb{R}^2$ be a compact $2$-dimensional submanifold, and let $\sigma\colon N\to \mathcal{S}$ be an almost Cartan section of the bundle $\pi_{\mathcal{S}}$. Suppose $w = X(x, y)\partial_x + Y(x, y)\partial_y$ is a vector field on $N$ such that
\begin{align*}
d\sigma(w|_p) \in \mathcal{C}_{\sigma(p)}\qquad \text{for all}\quad p \in N\,.
\end{align*}
Then, for $p = (x, y)\in N$, we have
\begin{align*}
d\sigma(w|_{p}) = X(x, y)\,\overline{\!D}_x|_{\sigma(p)} + Y(x, y)\,\overline{\!D}_y|_{\sigma(p)}.
\end{align*}
Assume that $X\neq 0$ in $N$. Without loss of generality, we can set $X = 1$. In this case, $\sigma$ has the following form
\begin{align*}
\sigma\colon\qquad\quad
\begin{aligned}
&u = f, \qquad u_x = \dfrac{\partial_x f + Y\partial_y f + Y\partial_x g + Y^2 \partial_y g}{1 + Y^2},\\
&v = g, \qquad v_x = -\dfrac{Y(\partial_x f + Y \partial_y f) - \partial_x g - Y\partial_y g}{1 + Y^2},\qquad
\end{aligned}
\ldots
\end{align*}
Here $f$ and $g$ are arbitrary functions on $N$. The expressions for all other coordinates on $\mathcal{S}$ are unambiguously defined. The pullback reads
\begin{align*}
&\sigma^*(\tilde{l}) = - \dfrac{(\partial_x f + Y\partial_y f)^2 - (\partial_x g + Y\partial_y g)^2 + 2(\partial_x g + Y\partial_y g)(Y\partial_x f - \partial_y f)}{2(1 + Y^2)}\,dx\wedge dy\,.
\end{align*}
Varying the corresponding action w.r.t. $f$, $g$ and $Y$, we get
\begin{align*}
&\partial_x \Big(\dfrac{\partial_x f + Y\partial_y f + Y\partial_x g + Y^2 \partial_y g}{1 + Y^2}\Big) + \partial_y \Big(\dfrac{Y(\partial_x f + Y \partial_y f) - \partial_x g - Y\partial_y g}{1 + Y^2}\Big) = 0\,,\\
&\partial_x \Big(\dfrac{Y\partial_x f - \partial_y f - \partial_x g  - Y\partial_y g}{1 + Y^2}\Big) + \partial_y \Big(\dfrac{Y(Y\partial_x f - \partial_y f - \partial_x g  - Y\partial_y g)}{1 + Y^2}\Big) = 0\,,\\
&(Y\partial_x f - \partial_y f - \partial_x g - Y\partial_y g)(\partial_x f + Y\partial_y f + Y\partial_x g - \partial_y g) = 0\,.
\end{align*}
The following two cases are due to the last equation.\\
\textbf{Case 1.}\quad $Y\partial_x f - \partial_y f - \partial_x g - Y\partial_y g = 0$.\quad In this case, we have 
$$
u = f,\qquad v = g,\qquad u_x = \partial_x f,\qquad v_x = -\partial_y f,\qquad \partial_x^2 f + \partial_y^2 f = 0\,.
$$
\textbf{Case 2.}\quad $\partial_x f + Y\partial_y f + Y\partial_x g - \partial_y g = 0$.\quad In this case, we have 
$$
u = f,\qquad v = g,\qquad u_x = \partial_y g,\qquad v_x = \partial_x g,\qquad -\partial_x^{\,2} g - \partial_y^{\,2} g = 0\,.
$$

Thus, we cannot exclude any sections from Case $1$, for which $f = 0$, as well as many other local sections. Moreover, one can show that the variation within the class of all almost Cartan embeddings also does not allow excluding them since $i_{\partial_v} d\tilde{l} = 0$. Let us note that $\partial_v$ is a symmetry of the Cauchy-Riemann equations, which is tangent to the fibers of the covering under consideration. 

So, the set of all stationary points of the internal Lagrangian contains more than just local solutions of $\pi_{\mathcal{S}}$.
}\\[-1ex]

\examplea{\label{potKdV} This example motivates the concept of $\xi$-stationary points.
Consider the potential KdV equation
\begin{align*}
u_t = 3u_x^2 + u_{xxx}
\end{align*}
and its infinite prolongation $\mathcal{E}$. Here $\pi\colon \mathbb{R}\times \mathbb{R}^2\to\mathbb{R}^2$ is the projection onto the second factor. Denote by $\,\overline{\!D}_t$ and $\,\overline{\!D}_x$ the restrictions of the total derivatives to $\mathcal{E}$. 

It is well known that this equation admits the symmetry with the characteristic $1 \in \ker l_{\mathcal{E}}$ and the presymplectic operator $\Delta = \,\overline{\!D}_x$ (see, e.g.,~\cite{Dorf}). As one can see, this characteristic belongs to the kernel of $\Delta$. The corresponding presymplectic structure $\omega$ is degenerate in this sense and can be related only to a variational principle that gives a consequence of the original equation, but not the potential KdV itself. This example is quite surprising because of that.

It is convenient to regard the variable $u_{xxx}$ and all its derivatives as external coordinates. Other coordinates on $J^{\infty}(\pi)$ can be treated as local coordinates on $\mathcal{E}$. There exists a unique internal Lagrangian $\boldsymbol\ell$ such that $\tilde{d}_1^{\,0,\,n-1} \boldsymbol\ell = \omega$. This internal Lagrangian is represented by the form
\begin{align*}
l = \Big(\dfrac{u_x u_t}{2} - u_x^3 + \dfrac{u_{xx}^2}{2}\Big)dt\wedge dx - \dfrac{1}{2}u_t\,dt\wedge \theta_0 + u_{xx}\,dt\wedge \theta_x + \dfrac{1}{2}u_x\, \theta_0\wedge dx\,,
\end{align*}
where $\theta_0 = du - u_x\, dx - u_t\, dt$ and $\theta_x = du_x - u_{xx}\, dx - u_{xt}\, dt$.

Let $N\subset \mathbb{R}^2$ be a compact $2$-dimensional submanifold, and let $\sigma\colon N\to \mathcal{E}$ be a $\xi$-section of the bundle $\pi_{\mathcal{E}}$ for $\xi = dx - X(t, x)dt$, where $X$ is a smooth function on $N$. Then $\sigma$ is of the form
\begin{align*}
\sigma\colon\qquad
\begin{aligned}
&u = f,\quad u_x = g,\quad u_{xx} = h,\quad u_t = \partial_t f + X(\partial_x f - g), \quad
u_{xt} = \partial_t g + X(\partial_x g - h),\quad \ldots
\end{aligned}
\end{align*}
Here $f$, $g$ and $h$ are arbitrary functions on $N$. The expressions for all other coordinates on $\mathcal{E}$ are unambiguously defined. We get
\begin{align*}
\sigma^*(l) = \Big(- \dfrac{\partial_x f \, \partial_t f}{2} + g\,\partial_t f - g^3 + h\,\partial_x g - \dfrac{h^2}{2} - \dfrac{X}{2}(\partial_x f - g)^2\Big)dt\wedge dx\,.
\end{align*}
Varying the corresponding action w.r.t. $f$, $g$ and $h$, we obtain
\begin{align*}
&\partial_t (\partial_x f - g) + \partial_x \big(X(\partial_x f - g)\big) = 0\,,\\
&\partial_t f - 3g^2 - \partial_x h + X(\partial_x f - g) = 0\,,\\
&\partial_x g - h = 0\,.
\end{align*}
These equations do not imply the relation $\partial_x f = g$. Therefore, the set of $\xi$-stationary points contains more than just local solutions. However, we also can vary w.r.t. $X$ (i.e., we can perturb $\ker \xi$). As a result, we get the missing equation
\begin{align*}
\partial_x f = g\,.
\end{align*}

Thus, if $\xi$ determines a non-characteristic distribution, then a $\xi$-section $\sigma$ is a stationary point of the internal Lagrangian $\boldsymbol\ell$ iff $\sigma$ is a local solution. But this is not the case for $\xi$-stationary points. 
}

\section{Example of degenerate Euler-Lagrange equations\label{degen}}

As an example of degenerate Euler-Lagrange equations, we consider the Proca equations for massive spin-$1$ field of mass $m$. For the sake of simplicity, we examine the Proca theory in Minkowski spacetime.
Let $\pi\colon \mathbb{R}^n\times \mathbb{R}^n \to \mathbb{R}^n$ be the projection onto the second factor, $n\geqslant 2$. We use the notation $t = x^{0}$, $x^1$, \ldots, $x^{n-1}$ for coordinates on the base space of $\pi$, while $A^0$, \ldots, $A^{n-1}$ denote coordinates along the fibers of $\pi$.

Consider the following differential $n$-form
\begin{align*}
L = \Big(\dfrac{1}{2}m^2A^\nu A_\nu - \dfrac{1}{4}F^{\mu\nu}F_{\mu\nu} \Big)d^nx\,,\qquad d^nx = dx^0\wedge \ldots \wedge dx^{n-1}\,.
\end{align*}
Here $F^{\mu\nu}$ denote $\partial^{\mu}A^{\nu} - \partial^{\nu}A^{\mu}$; the metric $\mathrm{diag} (+1, -1, \ldots, -1)$ is used to raise and lower indices. We assume that the indices $\mu$ and $\nu$ can take all the values $0$, \ldots, $n-1$, while the indices $i$, $j$ can take only the spatial values: $i, j = 1, \ldots, n-1$. 

The form $L$ is degenerate since the corresponding Euler-Lagrange equations
\begin{align}
\begin{aligned}
\partial_{\mu}F^{\mu\nu} + m^2A^{\nu} = 0
\end{aligned}
\label{Proc}
\end{align}
admit a lower order differential consequence. More specifically, applying $\partial_{\nu}$ to~\eqref{Proc} leads to the relation $\partial_{\nu}A^{\nu} = 0$. The corresponding internal Lagrangian is given by the restriction of the differential form
\begin{align*}
L + \omega_L = \Big(\dfrac{1}{2}m^2A^\nu A_\nu - \dfrac{1}{4}F^{\mu\nu}F_{\mu\nu} \Big)d^nx - F_{\mu\nu}\theta^\nu\wedge (i_{\partial^\mu}\, d^nx)
\end{align*}
to the infinite prolongation $\mathcal{E}$ of the Proca equations.

We can treat $x^{\mu}$, $A^i$, $F^{0i}$ and the spatial derivatives of $A^i$ and $F^{0i}$ as local coordinates on $\mathcal{E}$. Indeed, suppose that $F^{0i}$ is not just a notation but additional dependent variables. The Proca system is equivalent to the following one
\begin{align*}
&\partial_{\mu}(\partial^{\mu}A^{\nu} - \partial^{\nu}A^{\mu}) + m^2 A^{\nu} = 0\,,\\
&F^{0i} = \partial^0 A^i - \partial^i A^0\,.
\end{align*}
The infinite prolongation of this system coincide with the infinite prolongation of
\begin{align}
\begin{aligned}
&\partial_0 F^{0i} + \partial_j(\partial^{j}A^{i} - \partial^{i}A^{j}) + m^2A^i = 0\,,\\
&m^2A^0 = \partial_j F^{0j}\,,\\
&F^{0i} = \partial^0 A^i - m^{-2}\partial^i\partial_j F^{0j}\,.
\end{aligned}
\label{almPro}
\end{align}
Finally,~\eqref{almPro} is equivalent to the following evolutionary system
\begin{align}
\begin{aligned}
&\partial_0 F^{0i} = - \partial_j(\partial^{j}A^{i} - \partial^{i}A^{j}) - m^2A^i\,,\\
&\partial_0 A^i = F^{0i} + m^{-2}\partial^i\partial_j F^{0j}\,.
\end{aligned}
\label{Proevol}
\end{align}
Then the restrictions of the total derivatives $D_0$, \ldots, $D_{n-1}$ to $\mathcal{E}$ have the form
\begin{align*}
&\,\overline{\!D}_0 = \partial_0 + (F^{0i} + m^{-2} \partial^i \partial_j F^{0j})\partial_{A^{i}} - (\partial_{j}\partial^{j}A^{i} - \partial_j\partial^i A^j + m^2 A^{i})\partial_{F^{0i}} + \ldots\,,\\
&\,\overline{\!D}_i = \partial_i + \partial_i A^{j}\partial_{A^{j}} + \partial_i F^{0j}\partial_{F^{0j}} + \ldots
\end{align*}

Let $N\subset \mathbb{R}^n$ be a compact $n$-dimensional submanifold. Any $dt$-section $\sigma\colon N\to \mathcal{E}$ has the form
\begin{align*}
\sigma\colon\qquad A^i = f^i\,,\qquad F^{0i} = g^i\,,\qquad \partial_j A^{i} = \partial_j f^{i}\,, \qquad
\partial_j F^{0i} = \partial_j g^i\,,\qquad \ldots
\end{align*}
Here we use the notation $\partial_j f^{i}$, $\partial_j g^{i}$ for the partial derivatives $\partial_{x^j} f^{i}$, $\partial_{x^j} g^{i}$, while $\partial_j A^{i}$, $\partial_j F^{0i}$ denote coordinates on $\mathcal{E}$. Arbitrary functions $f^{i}, g^i\in C^{\infty}(N)$ determine a $dt$-section and vice versa. Since $A^0|_{\mathcal{E}} = m^{-2}\partial_i F^{0i}$, we get
\begin{align*}
\sigma^*(l) = \Big(\dfrac{1}{2}m^2f^i f_i + \dfrac{1}{2m^2}(\partial_ig^i)^2 + \dfrac{1}{m^2}g^j \partial_j\partial_i g^i + \dfrac{1}{2}g^ig_i - \dfrac{1}{4}(\partial^i f^j - \partial^j f^i)(\partial_i f_j - \partial_j f_i)  - g_i \partial_0 f^i \Big)d^nx\,,
\end{align*}
where $l = (L + \omega_L)|_{\mathcal{E}}$. The corresponding Euler-Lagrange equations are
\begin{align*}
&\partial_0 g^i = - \partial_j(\partial^j f^i - \partial^i f^j) - m^2 f^i\,,\\
&\partial_0 f^i = g^i + m^{-2}\partial^i \partial_j g^j\,.
\end{align*}

Comparing these Euler-Lagrange equations with~\eqref{Proevol}, we can conclude that 
all the $dt$-stationary points of the internal Lagrangian are solutions to the Proca equations.
This is probably due to the fact that the Proca equations can be written in an evolutionary form. Note that the evolutionary equations we have obtained are variational. Actually, the corresponding Lagrangian is given by the expression for the $\sigma^*(l)$.

\section{Conclusion}

Stationary action principles produce internal variational principles. The description of internal variational principles in terms of internal Lagrangians seems reasonable for infinitely prolonged systems of equations that can be written in an extended Kovalevskaya form, at least in a neighborhood of any point. In this case, provided that the space+time decomposition is fixed (or that the dimension of the Cartan distribution is less than or equal to two), we can restrict ourselves to considering unconstrained variational problems.

The role of perturbations of space+time decompositions is unclear to the author. Nevertheless, the concept of stationary points of an internal Lagrangian is remarkable in that it is invariant. Indeed, it is not based on any space+time decompositions, does not require a bundle structure on a differential equation, and does not depend on the choice of a particular representative of an internal Lagrangian. Apparently, in some cases, we somehow need to eliminate almost Cartan submanifolds composed of characteristic surfaces. Therefore, we need to define characteristics in terms of the intrinsic geometry of infinitely prolonged equations.

In the case of gauge systems, a modification of the concept of stationary points of an internal Lagrangian is required. Internal Lagrangians perhaps should be redefined using gauge distributions in this case.

\vspace{3.0ex}

\centerline{\bf{\Large Acknowledgments}}

\vspace{2.0ex}

The author is grateful to M.~Grigoriev for illuminating discussions. 
The author also appreciates significant discussions with I.S.~Krasil'shchik, and A.M.~Verbovetsky.

\end{document}